# Direct imaging and control of Berry curvature in noncollinear antiferromagnetic single-crystal thin films

Yuchuan Yao,[1,†] Pratap Pal,[1,†] Camron Farhang,[2,†] Weihang Lu,[2,†] Mohamed Elekhtiar,[3] Paul Lenharth,[1] Neil G. Campbell,[4] Gautam Gurung,[5,6] Roger D. Johnson,[7,8] Pascal Manuel,[9] Mark S. Rzchowski,[4] Evgeny Y. Tsymbal,[3] Jing Xia,[2*] and Chang-Beom Eom[1*]

[1]Department of Materials Science and Engineering, University of Wisconsin-Madison, Wisconsin 53706, USA
[2]Department of Physics and Astronomy, University of California, Irvine, CA 92696, USA
[3]Department of Physics and Astronomy & Nebraska Center for Materials and Nanoscience, University of Nebraska, Lincoln, NE 68588, USA
[4]Department of Physics, University of Wisconsin-Madison, Wisconsin 53706, USA
[5]Clarendon Laboratory, Department of Physics, University of Oxford, Parks Road, Oxford OX1 3PU, UK
[6]Trinity College, University of Oxford, Oxford OX1 3BH, UK
[7]Department of Physics and Astronomy, University College London, London, WC1E 6BT, UK
[8]London Centre for Nanotechnology, University College London, London WC1E 6BT, UK
[9]ISIS Facility, STFC Rutherford Appleton Laboratory, Didcot, Oxfordshire OX11 0QX, UK

*Corresponding authors: xia.jing@uci.edu, ceom@wisc.edu
†These authors contributed equally.

The discovery of the intrinsic anomalous Hall effect (AHE) in noncollinear antiferromagnets where transverse Hall voltage emerges without magnetic field, has opened a plethora of promising opportunities in antiferromagnetic devices. The key challenges limiting their full potential are (1) high-quality epitaxial thin-film growth and (2) the understanding of Berry curvature domain physics. Here, we focus on a noncollinear antiperovskite antiferromagnet $Mn_3NiN$ as a model system, successfully grown as a single-crystal epitaxial thin film. Combining multiple experiments supported by theoretical calculations, we probe the Berry curvature associated with antiferromagnetic $\Gamma_{4g}$ domains in $Mn_3NiN$ and its strong connection to an AHE. We directly image the intrinsic Berry curvature with high-resolution Sagnac microscopy, controlling spatial distribution and dynamics by varying temperature and applied magnetic fields. We discover that the $\Gamma_{4g}$ domains are switchable near the Néel transition, but become frozen and unresponsive to external stimuli at low temperature. This behavior enables the tuning of Berry-curvature driven AHE and magneto-optic Kerr effect responses through controlled experimental conditions. Our findings provide critical advancement of the fundamental understanding and wide tunability of Berry curvature in noncollinear antiferromagnets important for realization in potential spintronic applications.

## Introduction

Understanding and manipulating the coupling of electronic spin and charge in antiferromagnets offers enormous potential applications with the advantages of lack of stray fields, low susceptibility to external perturbations, and ultrafast spin dynamics (*1–10*). However, the absence of net magnetization presents significant challenges for the detection and control of the antiferromagnetic (AFM) state (*3*). Recent progress has come from the discovery of a large anomalous Hall effect (AHE) in noncollinear antiferromagnets (NCAFs) such as Mn based hexagonal $Mn_3Sn$ (*11*) or $Mn_3Ge$ (*12*), cubic $Mn_3Pt$ (*13*) or $Mn_3Ir$ (*14*) and antiperovskites $Mn_3XN$ ($X$=Ga (*15*), Sn (*16*) and Ni (*17*)). These NCAFs have nearly compensated, or a very weak spin-orbit coupled driven tilted moment in or out of the (111) Kagome plane (*10*, *11*), thus precluding a simple origin of the AHE in these systems driven by the net magnetic moment. Whereas recent theoretical studies have highlighted the role of spin-orbit-coupling with the simultaneous absence of symmetries (e.g. mirror symmetries in $\Gamma_{4g}$ phase of $Mn_3NiN$), in making the sum of Berry curvature over the Brillouin zone non-zero (*18*, *19*), resulting in a nonzero AHE. Thus, it raises a critical question of what drives the AHE in such NCAFs and how to control it like conventional ferromagnetic materials for information writing.

While the information writing/reading via controlling magnetic domains is well established in conventional ferromagnets (*20*), such control in an antiferromagnet is extremely challenging. However, some recent progress has shown promise for the non-collinear antiferromagnet ($Mn_3Sn$), where magnetic domains were mapped via magneto-optical-Kerr-effect (MOKE) imaging (*21*) in a bulk sample. But the hexagonal structure and complex magnetic configuration of $Mn_3Sn$ not only complicates understanding but also brings significant challenges for epitaxial thin-film growth (*11*). Use of NCAF materials with large AHE for spintronics requires epitaxial growth of these materials with simple crystal and spin structures. This not only provides better understanding, but also offers simpler integration with other functional materials in a device heterostructure.

Here, we choose $Mn_3NiN$ with antiperovskite structure as a model system, which has a simple cubic crystal structure with lattice parameter close to commercially available substrates and $\Gamma_{4g}$ triangular spin configuration(*22*). Such a spin configuration breaks time-reversal symmetry, resulting in a non-vanishing Berry curvature, as found in our first-principles density-functional-theory (DFT) calculations (fig. S1). Interestingly, $Mn_3NiN$ hosts two different kinds of magnetic domains ($D$ and $D'$) which give rise to positive and negative AHE (Fig. 1B), originating from opposite chiralities of the spin configurations. Such binary response of AHE, despite the antiferromagnetic nature, makes $Mn_3NiN$ a very compelling model system for information writing in a simple '0' and '1' scheme (*23*). Moreover, $Mn_3NiN$ is an interesting and promising material platform not only for fundamental investigations into the nature of the Berry curvature domains, but also for achieving efficient and deterministic magnetic domain writing. To overcome the limited spatial resolution of conventional MOKE, we utilize a zero-loop Sagnac interferometer with unprecedented MOKE sensitivity (*24*) to directly image smaller NCAF $\Gamma_{4g}$ magnetic domains responsible for berry phase AHE in $Mn_3NiN$ thin film. This provides direct evidence of unusual temperature dependent AHE in $Mn_3NiN$ thin film. Combining our temperature-dependent transport measurements and theoretical calculations, we provide a comprehensive framework for spintronic applications.

We demonstrate the growth of high-quality atomically smooth epitaxial $Mn_3NiN$ thin films, establish their robust antiferromagnetic properties, and observe a large AHE, that is strongly correlated with the underlying nanoscale Berry curvature domain texture. Our films show noncollinear $\Gamma_{4g}$ magnetic ordering below the Néel temperature ($T_N$) of 240 K that retains this symmetry to the lowest measured temperature of 2 K. Importantly, we observe a large AHE ($\rho_{xy}$ ~1 μΩ-cm) around $T_N$ comparable in size to common ferromagnets, even though the films show a net moment of only 0.008 $\mu_B$/Mn (arising from a small distortion of the non-collinear $\Gamma_{4g}$-spin arrangement in the (111) plane). This large AHE in the absence of a significant net moment suggests a Berry phase mechanism, and the weak moment provides a controllability to switch AFM domains, and hence the AHE, with applied magnetic field. Furthermore, we provide direct MOKE imaging and deterministic controllability of the $\Gamma_{4g}$ magnetic domains giving rise to large berry curvature driven AHE, via a zero-loop Sagnac interferometer. Our findings reveal that Berry curvature domains could be switched and saturated with an applied magnetic field of 9 T just below Néel temperature, whereas they become unswitchable at low temperature, which could be explained by magnetic anisotropy, as schematically illustrated in Fig. 1C. However, we can still use the field cooling to align the Berry curvature domains, resulting in saturated AHE at low temperature, indicating our precise controllability of berry curvature domains, which are further proved by theory calculations. These new insights advance our understanding of the AHE order in NCAFs and will provide crucial guidance for harnessing AHE in AFM-based electronics.

## Results and Discussions

### Epitaxial $Mn_3NiN$ thin film with robust Magnetic phase transition

$Mn_3NiN$ is an antiperovskite with cations (Mn) and anions (N) positions interchanged compared to typical $ABO_3$ perovskite. Whereas $ABO_3$ perovskites can be viewed as alternate stackings of $A$O and $B$O$_2$ layers along the [001] direction, $Mn_3NiN$ can be described as similar stackings of MnNi and $Mn_2N$ layers (fig. S2). Importantly, we were able to grow high quality $Mn_3NiN$ thin films on $(La_{0.3}Sr_{0.7})(Al_{0.65}Ta_{0.35})O_3$ (LSAT) substrates using optimized growth conditions in a dc reactive magnetron sputtering system, guided by our previous research (*9*, *15*). Out-of-plane X-ray diffraction shows a strong $Mn_3NiN$ peak with distinct Kiessig fringes, indicating high crystalline quality that is further corroborated by the streaky RHEED pattern (figs. S3 and S4). X-ray reciprocal space mapping (RSM) measurements centered on the asymmetrical ($\bar{1}13$) peak (fig. S3B) indicate that *c/a*=0.9952 at room-temperature. Atomic force microscope imaging shows an atomically smooth surface with a roughness of ~0.3 nm (fig. S3C). These data (further details are provided in figs. S4-S6) establish the high-quality epitaxial growth of single-crystal $Mn_3NiN$ thin films, allowing conclusive magnetic, transport, and optical measurements.

To investigate the transport properties, we fabricated Hall bar devices on such high-quality $Mn_3NiN$ epitaxial film (inset to Fig. 2A). The longitudinal resistivity shows a clear kink at the paramagnetic to antiferromagnetic Néel transition of 240 K (Fig. 2A), consistent with our magnetization measurements (fig. S7). The transverse resistivity shows a transition from a linear dependence on magnetic field above $T_N$ consistent with an ordinary Hall effect, to a highly nonlinear field dependence with a clear hysteresis below $T_N$ (Fig. 2B), indicating contribution from a nontrivial AHE (*11*). We extracted the AHE component by subtracting the high-field linear part from the total Hall signal (fig. S10A) which shows that the resulting AHE component has a hysteretic behavior at low applied magnetic field, and saturation at high magnetic field. The high-field saturation value arises from AFM domain alignment. We deduced anomalous Hall conductivity (AHC) from anomalous Hall resistivity (AHR) and sheet resistivity data as $\sigma_{xy} = -\rho_{xy}/\rho_{xx}^2$.

Interestingly, the $\sigma_{xy}$ of a 40 nm $Mn_3NiN$ thin film exhibits an unusual temperature dependence with a maximum value of about ~15 $\Omega^{-1}$-cm$^{-1}$ (Fig. 2C) and anomalous Hall resistivity $\rho_{xy}$~1 μΩ-cm (fig. S10). This behavior was independent of thickness as a similar result was obtained on a 20 nm thin film, excluding any role of the surface magnetization contribution (fig. S11) (*25*). The large intrinsic value of anomalous Hall is of immense interest and importance, as it is comparable to common ferromagnetic metals such as 0.6 μΩ-cm for Fe with net moment 2.18 $\mu_B$/Fe, 0.1 μΩ-cm for Co with moment 1.67 $\mu_B$/Co, and 0.15 μΩ-cm for Ni with moment 0.62 $\mu_B$/Ni (*26–30*). The small 0.008 $\mu_B$/Mn magnetic moment of our $Mn_3NiN$ films (fig. S7), is too small to generate large $\sigma_{xy}$ and $\rho_{xy}$ (*31*) in our $Mn_3NiN$ thin films like conventional mechanisms in ferromagnets. This large $\sigma_{xy}$ and $\rho_{xy}$ of $Mn_3NiN$ films is comparable to that found in other non-collinear antiferromagnets (fig. S12), which is consistent with a non-vanishing Berry curvature from the $\Gamma_{4g}$ spin structure, capable of inducing such a large AHE (*14*, *32*, *33*). Thus, we conclude that the effect of the net magnetic moment is secondary in generating the observed AHE. Comparing the anomalous behavior of $\sigma_{xy}$ with the net moment (fig. S13), further supports this conclusion.

The temperature dependent response of anomalous $\sigma_{xy}$ exhibits non-trivial trend. While the $\sigma_{xy}$ is zero above $T_N$, it increases to large values (~15 $\Omega^{-1}$-$cm^{-1}$) immediately below $T_N$, before gradually dropping close to zero as temperature is further decreased below ~100 K. One possible explanation for the unusual disappearance of AHE at low temperatures is the absence of finite Berry curvature, potentially due to a magnetic phase transition from $\Gamma_{4g}$ to $\Gamma_{5g}$ states, similar to what has been observed in case of $Mn_3Pt$ (*13*). However, this scenario is rather exotic and contradicts our observations of robust $\Gamma_{4g}$ antiferromagnetic ordering even down to 2 K, as found in our neutron diffraction measurements (fig. S7 and S8). Another possibility could be the formation of Berry curvature domains (*21*), where equal population of domains of opposite polarity can cancel, leading to zero $\sigma_{xy}$ from the entire sample. Thus, to have a holistic understanding of the origin of such unique temperature dependence of anomalous Hall, the direct observation of Berry curvature domains dynamics is of high importance to advance the research on emergent non-collinear antiferromagnetic spintronics.

**Direct Sagnac MOKE imaging of Berry Curvature domains**

MOKE imaging can directly reveal Berry curvature domain structures, as previously demonstrated in $Mn_3Sn$ (*21*), where the Kerr rotation angle $\theta_k$ serves as a local probe of $\sigma_{xy}$ at optical frequencies (Fig. 3A). In this work, we employ a scanning fiber-optic Sagnac interferometer, capable of detecting Kerr rotation with high resolution nano-radian sensitivity (fig. S14), integrated into a scanning optical probe for micron-resolution imaging of Kerr rotation, as described in previous publications (*24*, *34*, *35*). Unlike conventional MOKE or AHE techniques, which often rely on magnetic hysteresis to isolate intrinsic signals, the Sagnac interferometer's time-reversal symmetric design inherently suppresses nonreciprocal artifacts such as optical misalignment and extrinsic scattering, enabling precise, field-stable measurements of $\theta_k$. We perform Sagnac MOKE imaging and single point measurements across a range of temperatures and magnetic fields (Fig. 3B and fig. S15A) on Hall bar devices (Fig. 3C), capturing the domain evolution and enabling direct correlation with transport signatures of Berry curvature.

We begin by examining the temperature dependence of the Kerr rotation $\theta_k$. Field-cooled (FC) measurements of $\theta_k$ at fixed points on the sample surface during a warm-up in zero field reveal a sharp onset of $\approx 60\ \mu rad$ Kerr rotation below $T_N$, which remains nearly constant down to 2 K (fig. S15A). This temperature-independent behavior of $\theta_k$, excellently corresponds to the magnetization and neutron data, however contrasts sharply with the strong temperature dependence of the anomalous Hall conductivity $\sigma_{xy}$ (Fig. 2C), suggesting that the underlying magnetic order remains stable while the transport response evolves.

To probe the temperature evolution of the spatial distribution of Berry curvature domains on the sample surface, we first repeat the process of cooling the sample under +0.3 T applied magnetic field to polarize all domains before removing the field and performing scanning images of $\theta_k$ in zero-field, both above and below $T_N$ (fig. S15B). At 300 K, a scan of $\theta_k$ shows a uniform and near zero value across the entire sample, consistent with the absence of net Berry curvature. In contrast, at 2 K, we observe a strong, spatially homogeneous Kerr response across the entire area, indicating a coherent alignment of Berry curvature domains. These results confirm the robustness of the $\Gamma_{4g}$ magnetic order at low temperatures. However, the mismatch between the static almost uniform Kerr response (fig. S15) and the evolving $\sigma_{xy}$ (Fig. 2) highlights that domain dynamics alone cannot fully account for the unusual temperature dependence of the AHE in this system.

To further investigate the role of Berry curvature domains and their switching behavior in $Mn_3NiN$, we performed magnetic field-dependent MOKE measurements at 250 K and 210 K (Fig. 3B), along with scanning at key points along the 210 K hysteresis loop, including positive and negative saturation fields (±9 T), the coercive field, and zero field (Fig. 3C). At 210 K, well below the magnetic transition, we observe a pronounced Kerr hysteresis loop that reaches a saturated $\theta_k$ of $\approx 200\ \mu rad$ below 9 T, on scale with the largest reported Kerr rotation in NCAFs (*21*). The field dependence at 250 K remains nearly linear. At 210 K, both the anomalous Hall conductivity and remnant $\theta_k$ reach magnitudes comparable to those of $Mn_3Sn$ (*21*). A remnant $\theta_k$ of $\approx 54\ \mu rad$ persists at zero field after sweeping back down from +9 T, indicating the equivalence of +0.3 T field cooling (fig. S15A), and fully polarizing the sample at 210 K with a 9T field.

Kerr imaging at selected field values during the 210 K hysteresis measurements reveals key insights into the switching mechanism. The switching of Berry curvature domain polarization proceeds through an apparently uniform evolution across the entire 50 $\mu$m by 150 $\mu$m area of the Hall bar (Fig. 3C). From -9 T to +9 T, the polarization transitions coherently between $\mp 200\ \mu rad$, without visible evidence of domain wall motion at this scale. The findings in Fig. 3C, indicate that the domain reversal mechanism in $Mn_3NiN$ is predominantly coherent over macroscopic length scales, but may involve fine-scale domain evolution at the mesoscopic or microscopic level. Therefore, we perform higher resolution scans in the following section to test for potential subtle spatial variations in $\theta_k$, that reveal the presence of weak domain textures not resolved in the coarse scans in Fig. 3C, and also try to rule out the inconsistent transport response at low temperature.

**Microscopic understanding of unconventional temperature dependent AHE**

We find that $Mn_3NiN$ exhibits a large AHC near the Néel temperature, which progressively diminishes upon further cooling and ultimately vanishes below 100 K. This behavior allows us to define two distinct regimes: Region I (100 K<T<240 K), where a significant AHC is observed, and Region II (T<100 K), where the AHC is essentially absent. To elucidate the microscopic origin of this unusual temperature dependence of $\sigma_{xy}$ (Fig. 2C), we employ Sagnac MOKE as a deterministic probe of Berry-curvature domains and their coupling to the underlying $\Gamma_{4g}$ magnetic domains. We cooled $Mn_3NiN$ thin films from high temperature through the $T_N$ to 20 K under both ZFC and field-cooled FC conditions, then measured both AHC and $\theta_k$ as a function of magnetic field (Fig. 4A).

First, we made Hall measurements at 20 K (region II) after cooling the sample at positive, zero and negative applied fields (Fig. 4B). Such a poling under an applied magnetic field leads to complete freezing of antiferromagnetic domains at low temperature and thus, the sign of anomalous Hall is interestingly not switchable (Fig. 4B), likely due to increased magnetic anisotropy. Strikingly, the ZFC Kerr hysteresis loop (Fig. 4C, black curve) closely mirrors the AHE behavior (Fig. 4B), exhibiting zero remnant $\theta_k$ under ZFC conditions. In contrast, the FC loop (Fig. 4C, red curve) shows a finite remnant $\theta_k$ at zero field, unchanged from the spontaneous Kerr signal acquired during the +0.3 T field cooling, indicating the field cooling can effectively align the Berry curvature domain beforehand. However, similar to the AHE measurements, the sign of the remnant ZFC signal cannot be reversed, even by sweeping field to ±9 T, indicating that the Berry curvature domains are frozen in and robust against switching at low temperatures. The unswitchable nature of the domains at low temperature behavior suggests that a net Berry curvature emerges only when domains are aligned during cooling, or by large magnetic fields at sufficiently high temperatures. Additionally, the spontaneous $\sigma_{xy}$ and $\theta_k$ signals remain stable for many hours without measurable decay (fig. S16), further verifying a robust domain configuration below the $T_N$, giving a potential for permanent antiferromagnetic storage devices. These observations demonstrate that at low temperatures (region II in Fig. 3), Berry curvature domains become pinned and cannot be reconfigured by magnetic field alone, explaining the near-zero AHE seen in ZFC conditions.

To directly visualize this frozen behavior, we performed scanning Kerr imaging at 20 K after cooling under various field conditions, fully utilizing the $\sim 1\ um$ resolution of the scanning Sagnac probe.

A $\theta_k$ scan taken after ZFC to 20 K (Fig. 4D), reveals a disordered domain landscape with spatially random but near-zero $\theta_k$ in the range of $\pm 3\ \mu rad$, consistent with a vanishing net Berry curvature composed of weak, locally uncompensated domains. Assuming a random formation of fully saturated domains in the ZFC state, we can estimate the domain size as: $d = \frac{\sigma w}{\theta_{sat}} = \frac{2\mu rad \times 1.5\ \mu m}{200\mu rad} = 15$ nm, where $\sigma$, $w$, $\theta_{sat}$ are the standard deviation, beam width, and saturated Kerr rotation from Fig. 3B, respectively. Notably, the ZFC $\theta_k$ scan images taken at 20 K, after cooling in a +0.3 T field (Fig. 4D), both show Berry curvature landscapes that are indistinguishable from each other. These higher-resolution scans reveal fine domain textures with a variation of $\pm 4\ \mu rad$ from the polarized $\theta_k$ value that were not resolved in the earlier, zoomed-out images collected during the hysteresis loop (Fig. 3C). In those coarse scans, the large color scale and broader spatial averaging masked subtle variations in $\theta_{\mathrm{k}}$, making the Berry curvature landscape appear uniformly polarized. Here, the more sensitive imaging exposes the underlying domain structure, confirming that the apparent coherence in the previous measurements arises from macroscopic averaging over a multi-domain state.

Together, this data provides a microscopic explanation for the suppressed $\sigma_{xy}$ at low temperature (region II in Fig. 3) under ZFC: the Berry curvature domains are present but randomly oriented canceling out the macroscopic signal. Additionally, the Kerr rotation measurements as a function of temperature reveal that $\theta_{\mathrm{k}}$ remains nearly constant below the $T_N$, suggesting that the underlying Berry curvature persists to the lowest measured temperatures, even as the AHE vanishes under ZFC. And the scan images Fig. 4D confirm that the same Berry curvature landscape probed in both the $\sigma_{xy}$ (Fig. 2B) and Kerr hysteresis loop measurements at 210 K (Fig. 2B and Fig. 3B), exist down to low temperatures regardless of FC. This behavior confirms that the loss of anomalous Hall response at low temperature arises not from the suppression of Berry curvature itself, but from the inability to reorient the frozen domains below 210 K with the available magnetic field strength.

The ability to stabilize and control domain polarization through field cooling across $T_N$ highlights the importance of thermal history and magnetic field in orienting the domains in $Mn_3NiN$. The high sensitivity of the zero-area-loop Sagnac interferometer is essential for detecting these weak, domain signals in zero field; without it, the presence of frozen Berry Curvature domains might have remained hidden.

**Theoretical modeling of temperature dependent domain switching**

To demonstrate the magnetic domain switching of $Mn_3NiN$ with the $\Gamma_{4g}$ spin ordering as a function of temperature, we perform first-principles and atomistic spin-dynamics simulations. We consider the exchange coupling between first and second nearest Mn atoms in the Heisenberg Hamiltonian (Figs. 5A, and 5B, and fig. S17). We examine two scenarios of spin ordering at 20 K and 150 K by calculating the out-of-plane component of net magnetic moment $M_{net\perp}$. The sign of this moment is uniquely linked to D and D′ types of $\Gamma_{4g}$ antiferromagnetic domains and its reversal implies domain switching. We find that while the calculated $M_{net\perp}$ shows clear switchability with applied magnetic field in the range from -8T to 8T at 150 K (Fig. 5C), but exhibits two distinct FC unswitchable responses at 20 K (Fig. 5D). It is found that the large single-ion magnetic anisotropy is the primary factor that leads to such robust freezing of antiferromagnetic domains at low temperatures. Such an observation unanimously supports both the transport and Sagnac imaging data that $\Gamma_{4g}$ domains can be aligned during field cooling, but cannot be switched by sweeping magnetic field at lower temperatures (<~100K). At 150K, the $\Gamma_{4g}$ spin configuration is weakly distorted by thermal fluctuations, resulting in an effectively lower anisotropy barrier between domains (Fig. 1C), therefore, $\Gamma_{4g}$ domains become switchable under sufficient magnet fields. Herein, the theory calculation firmly supports the experimental observation of unusual temperature dependent switching domains in $Mn_3NiN$.

## Conclusions and outlook

We have successfully demonstrated the growth of high-quality epitaxial antiperovskite $Mn_3NiN$ single-crystal thin films on perovskite LSAT substates with noncollinear antiferromagnetic $\Gamma_{4g}$ phase and shown that its Berry-phase driven AHE ( $\sigma_{xy}$ ) through conventional magnetic and transport characterizations. Further, through Sagnac MOKE measurement, we directly imaged and precisely controlled the Berry curvature domains and AHE in $Mn_3NiN$ by temperature and magnetic field, explained by theoretical calculations. These results not only provided insight into the Berry curvature nature of $Mn_3NiN$, but also explained the unusual temperature-dependent AHE behavior, revealing a path toward new AHE-based spintronic systems. Looking beyond $Mn_3NiN$, it would be valuable to investigate other potential noncollinear antiferromagnetic systems using a Sagnac microscope, where current-induced switching of AHE and antiferromagnetic tunnel junction (AFMTJ) (*7*, *8*) showed significant promise toward spintronic applications. As the effective MOKE imaging intensively provides nanoscale probing of Berry curvature domain control in these noncollinear antiferromagnets, experimentally testing of these scenarios would be an interesting direction for future research.

**Acknowledgement**

C.B.E. acknowledges support for this research through a Vannevar Bush Faculty Fellowship (ONR N00014-20-1-2844), and the Gordon and Betty Moore Foundation's EPiQS Initiative, Grant GBMF9065. J.X. acknowledges support for this research through NSF award DMR-2419425 and the Gordon and Betty Moore Foundation EPiQS Initiative, Grant # GBMF10276. Transport measurement at the University of Wisconsin–Madison was supported by the US Department of Energy (DOE), Office of Science, Office of Basic Energy Sciences (BES), under award number DE-FG02-06ER46327. Theoretical modeling at the University of Nebraska-Lincoln (UNL) was supported by the National Science Foundation (grant No. DMR-2425567) and UNL's Grand Challenges catalyst award "Quantum Approaches Addressing Global Threats."

**Author Contributions**

Y.Y. P.P. and C.B.E. conceived the project. C.B.E. and J.X. supervised the project. Y.Y., P.P., and P.L. fabricated the samples and performed transport and magnetometry measurements. C.F. and W.L. performed Sagnac MOKE measurements. J.X. interpreted MOKE data. M.E., G.G., and E.Y.T. performed the theoretical calculations. N.C. and M.S.R. helped in measurements and analyses of the magnetic and transport data. R.D.J. performed the neutron diffraction measurements and analyses. Y.Y., P.P. J.X., and C.B.E. wrote the paper with inputs from all authors.

**Competing interests**
The authors declare no competing interest.

**Materials & Correspondence**
Correspondence and requests for materials should be addressed to Jing Xia (xia.jing@uci.edu) or Chang-Beom Eom (ceom@wisc.edu).

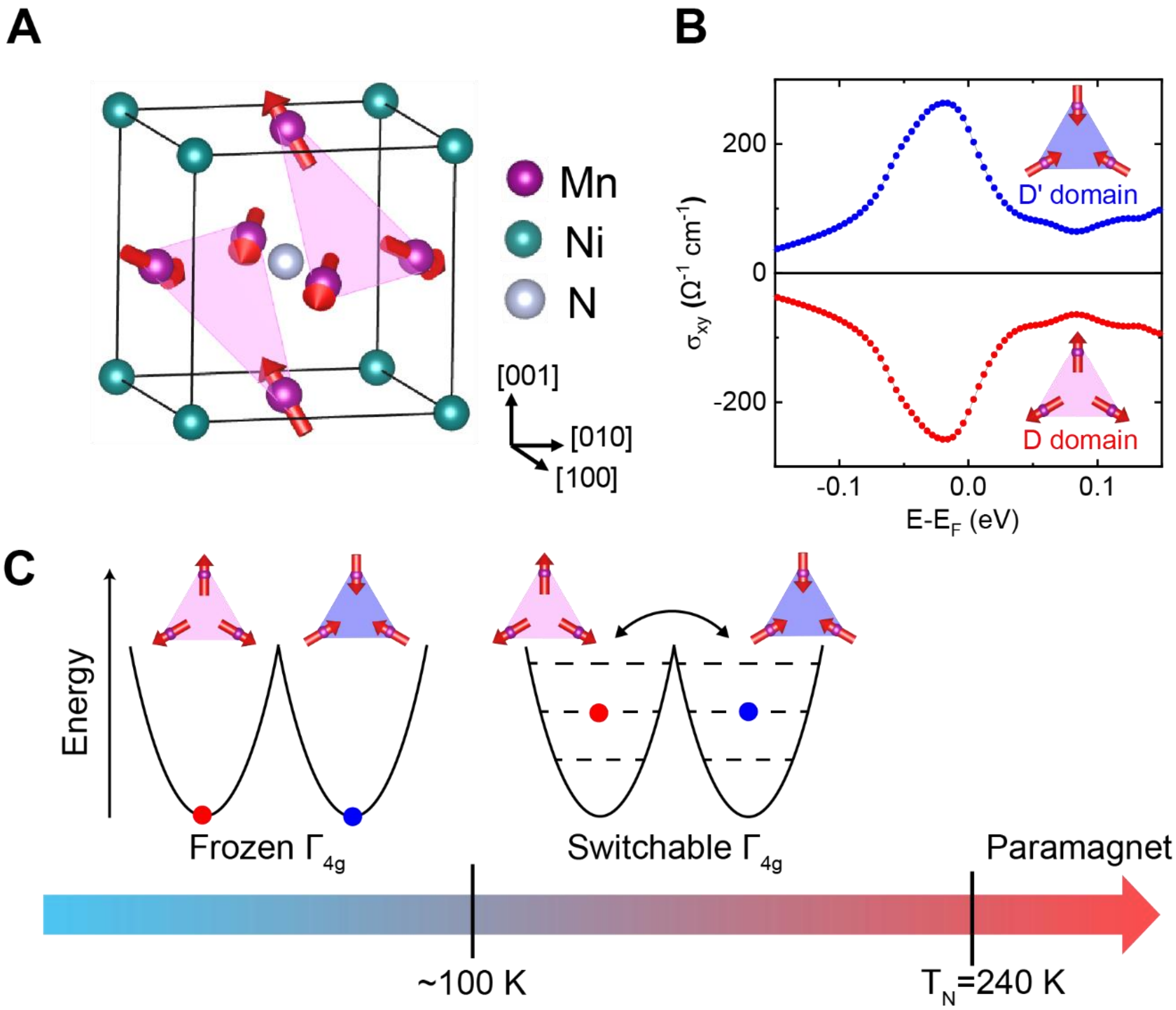

**Fig 1: Schematic illustration of unusual temperature dependent $\Gamma_{4g}$ spin order in $Mn_3NiN$. (A)** Unit cell crystal structure of Antiperovskite $Mn_3NiN$ with $\Gamma_{4g}$ spin configuration. **(B)** Two different types of magnetic domins in $Mn_3NiN$: *D* where spins point out and D′ where spins point in. Corresponding anomalous Hall conductivity for these two types of $\Gamma_{4g}$ spin configurations in $Mn_3NiN$, as deduced by DFT calculations. **(C)** Unusual temperature dependent response in $Mn_3NiN$ thin films due to strong anisotropy of magnetic domains.

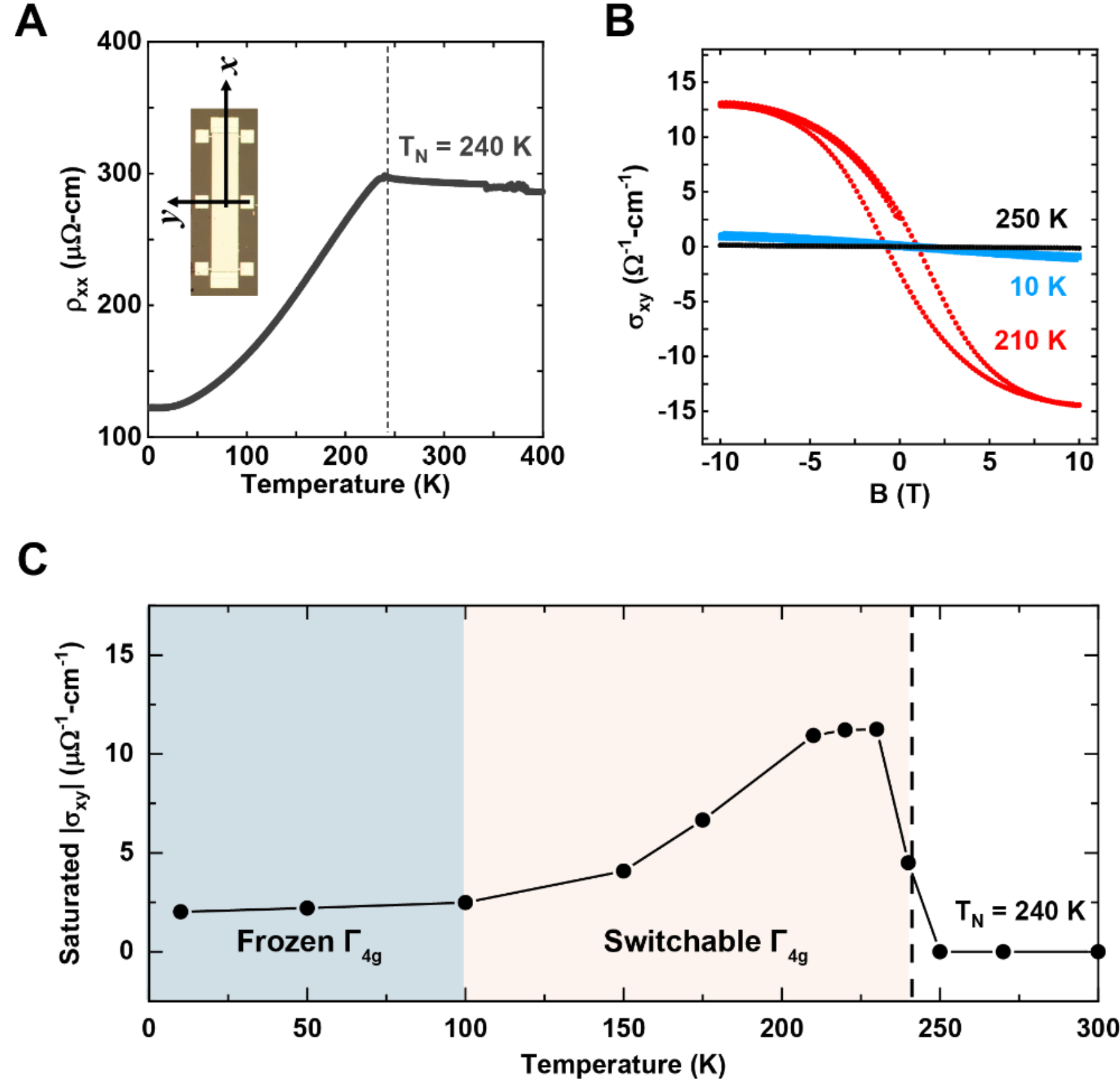

**Fig 2: Large anomalous Hall effect in $Mn_3NiN$ single-crystal thin films. (A)** Temperature dependent variation of longitudinal sheet resistivity showing excellent metallic behavior below the Néel transition of 240 K. Corresponding optical image of the Hall bar shown in the inset. **(B)** Hall conductivity vs out-of-plane applied magnetic field showing clear anomalous Hall signal below the $T_N$. **(C)** Temperature dependence of the saturated anomalous Hall conductivity change, showing maximum around the paramagnetic to $\Gamma_{4g}$ transition temperature and then gradually decreases and becomes almost zero below ~100 K.

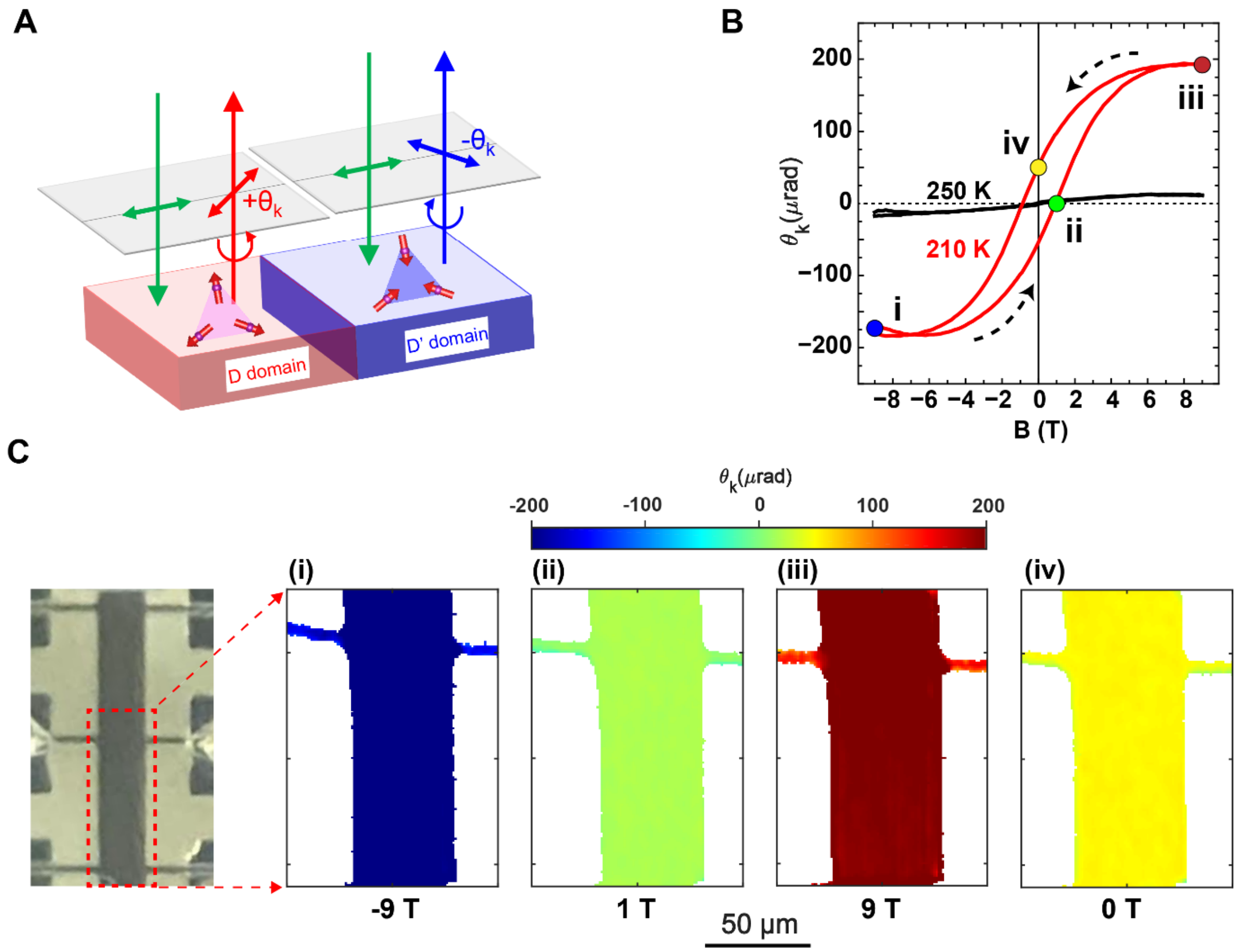

**Fig 3: Direct Sagnac MOKE imaging of magnetic domain switching in $Mn_3NiN$ thin films. (A)** Schematic illustration of magneto-optical response corresponding to different types of $\Gamma_{4g}$ domains in $Mn_3NiN$, leading to positive or negative Kerr rotations. **(B)** Kerr rotation $\theta_k$ vs out-of-plane applied magnetic field at 250 K (above $T_N$) and 210 K (below $T_N$), showing a pronounced hysteresis loop at the latter temperature with a saturated Kerr signal of $\theta_k \approx 200\ \mu rad$, and remnant $\theta_k$similar in magnitude to the value in Extended data Fig. 4a, with a nearly flat linear response at 250 K. **(C)** Scanning Kerr imaging taken at multiple field values during a hysteresis magnetic field sweep, with labels "i-iv" corresponding to the points marked in the hysteresis loop in panel **(B)**.

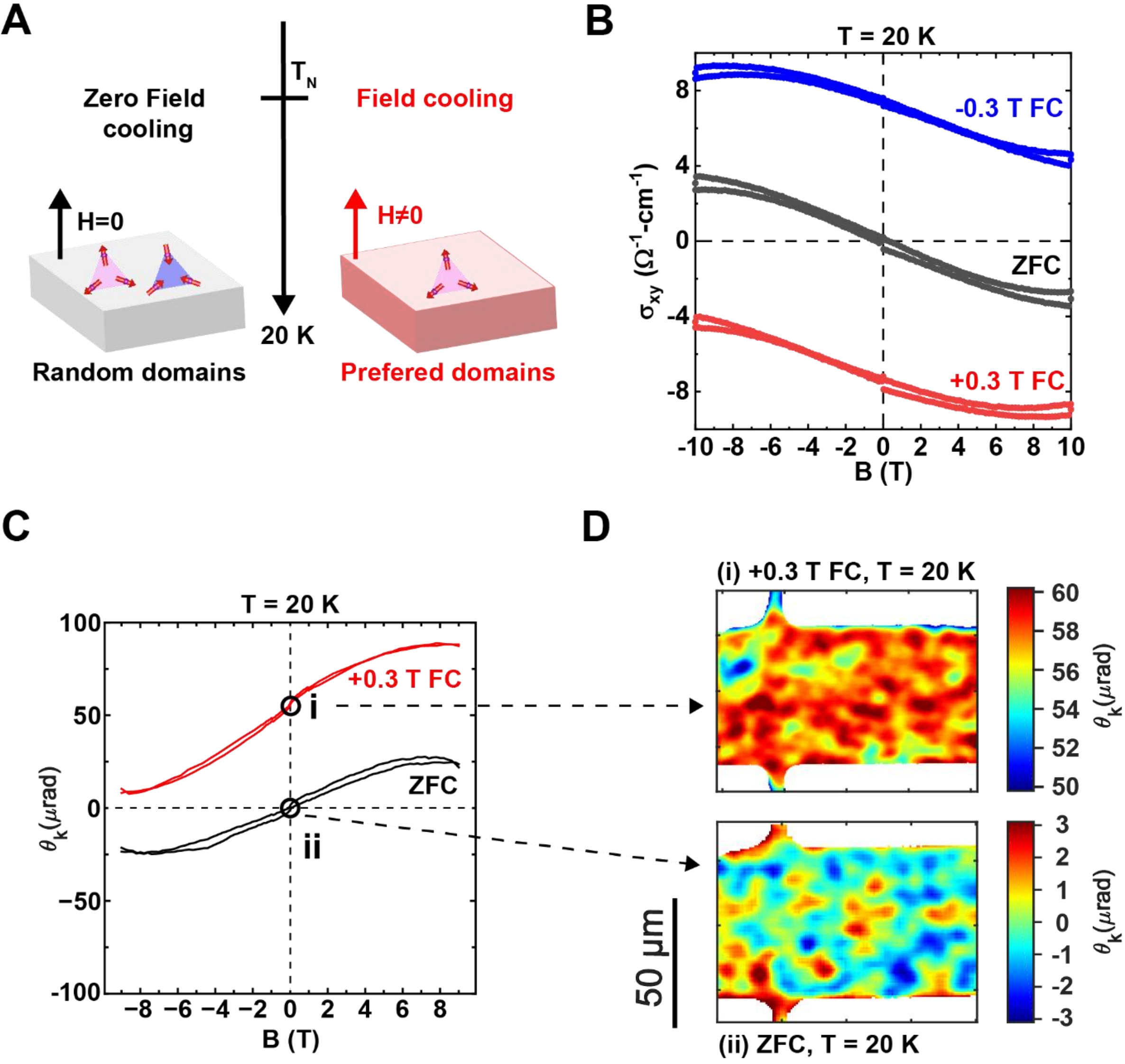

**Fig 4: Direct control of Berry curvature by applied poling magnetic field. (A)** Schematic illustration of the antiferromagnetic $\Gamma_{4g}$ domain alignment under zero or finite out-of-plane magnetic field cooling conditions in $Mn_3NiN$ thin film. **(B)** Variation of the anomalous Hall conductivity with magnetic field while ramping the field from 0 T to 10 T at 20K, after cooling the sample under zero field (ZFC) and ± 0.3 T. **(C)** Kerr rotation $\theta_k$ as a function of magnetic field under the same cooling conditions as in **a**, directly measured using a Sagnac interferometer. The ZFC curve exhibits a symmetric hysteresis with no remnant $\theta_k$, while the +0.3 T FC curve shows a robust remanent Kerr signal at zero field, indicating strong domain alignment. **(D)** Corresponding Scanning Kerr images at 0 T following ZFC and 0.3 T field cooling, revealing a disordered domain landscape with spatially random $\theta_k$ contrast centered around zero (±3 μrad) and a strongly polarized domain configuration with large, positive $\theta_k$ across the device respectively.

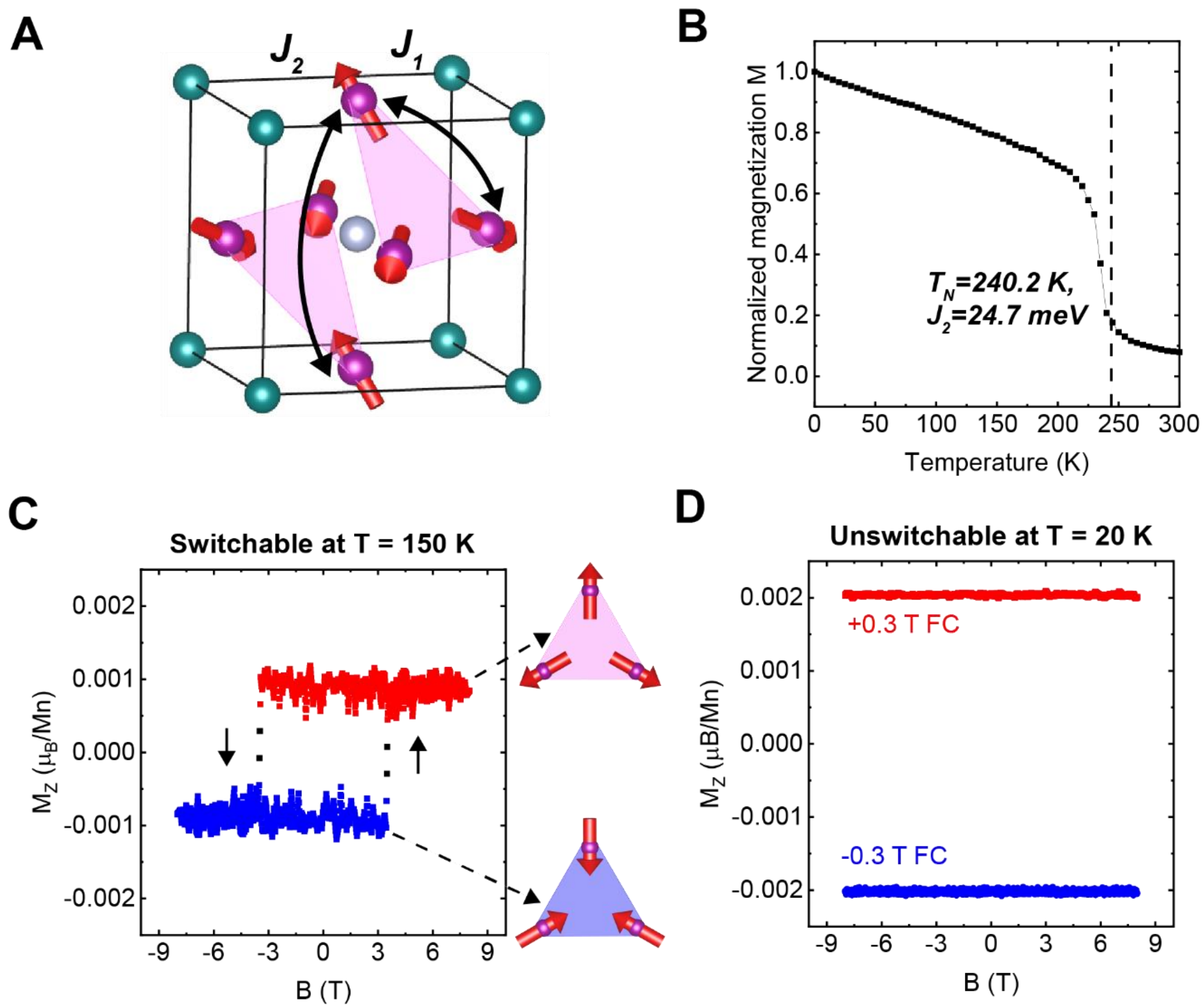

**Fig 5: Understanding of the magnetic domain dynamics via theoretical simulations of $Mn_3NiN$ thin films. (A)** Schematic illustration of the nearest and next nearest neighbor exchange coupling considered for calculations in $Mn_3NiN$. **(B)** Simulation of temperature dependent response of bulk magnetization, clearly showing the Néel transition at ~240 K, which is in excellent agreement with the experimental observations. **(C)** Corresponding calculation of the out-of-plane net magnetization switching vs applied magnetic field at 150K, where the $\Gamma_{4g}$ domain could be easily aligned and switched with $\Gamma_{4g}$ spin configuration. **(D)** Simulations of the out-of-plane net magnetization at 20K after different field cooling, displaying strong non-switchability of the $\Gamma_{4g}$ magnetic domains, agreeing with the experimental data. Domains can only be poled to write information by applied field from a temperature above $T_N$.

Supplemental Material for

# Direct imaging and control of Berry curvature in noncollinear antiferromagnetic single-crystal thin films

Yuchuan Yao,[1,†] Pratap Pal,[1,†] Camron Farhang,[2,†] Weihang Lu,[2,†] Mohamed Elekhtiar,[3] Paul Lenharth,[1] Neil G. Campbell,[4] Gautam Gurung,[5,6] Roger D. Johnson,[7,8] Pascal Manuel,[9] Mark S. Rzchowski,[4] Evgeny Y. Tsymbal,[3] Jing Xia,[2*] and Chang-Beom Eom[1*]

[1]Department of Materials Science and Engineering, University of Wisconsin-Madison, Wisconsin 53706, USA
[2]Department of Physics and Astronomy, University of California, Irvine, CA 92696, USA
[3]Department of Physics and Astronomy & Nebraska Center for Materials and Nanoscience, University of Nebraska, Lincoln, NE 68588, USA
[4]Department of Physics, University of Wisconsin-Madison, Wisconsin 53706, USA
[5]Clarendon Laboratory, Department of Physics, University of Oxford, Parks Road, Oxford OX1 3PU, UK
[6]Trinity College, University of Oxford, Oxford OX1 3BH, UK
[7]Department of Physics and Astronomy, University College London, London, WC1E 6BT, United Kingdom
[8]London Centre for Nanotechnology, University College London, London WC1E 6BT, United Kingdom.
[9]ISIS Facility, STFC Rutherford Appleton Laboratory, Didcot, Oxfordshire OX11 0QX, United Kingdom.

*Corresponding authors: xia.jing@uci.edu, ceom@wisc.edu
†These authors contributed equally.

**Supplementary Text 1**

**Materials and Methods**

**Thin-film growth and characterizations:** Epitaxial 40 nm $Mn_3NiN$ single-crystal thin films were grown on (001)-oriented $(La_{0.3}Sr_{0.7})(Al_{0.65}Ta_{0.35})O_3$ (LSAT) substrates at 650℃ under an $Ar/N_2$ (78%:22%) mixed atmosphere of 10 mTorr by DC reactive magnetron sputtering using a stoichiometric $Mn_3Ni$ target, which was monitored by in-situ reflection high energy electron diffraction (RHEED). $Mn_3NiN$ films were measured by atomic force microscopy and X-ray diffraction (XRD) spectroscopy for structural characterization.

**Magnetic and transport measurements:** DC magnetic measurements were carried out in a SQUID magnetometer on a 3*3 mm sample, and all magneto transport measurements were conducted in a Physical Property Measurement System on patterned Hall bar (50 μm × 200 μm) with Ti/Pt electrodes (first Ti was deposited which acts as a binder to the subsequent Pt electrodes).

**Neutron diffraction:** Single crystal neutron diffraction measurements were performed on the WISH time-of-flight diffractometer38 at ISIS, the UK neutron and muon source. An approximately 250 nm thick (001) $Mn_3NiN$ film samples with lateral dimensions 10 × 8 mm, were oriented for the measurement of nuclear and magnetic diffraction intensities in the (HK0) reciprocal lattice plane. The sample was mounted within a $^4He$ cryostat, and diffraction patterns were collected from a base temperature of 1.5 K up to 250 K.

**Sagnac MOKE measurements:** Magneto optical Kerr effect measurements were performed using a zero-loop fiber-optic Sagnac interferometer and microscope as described in our earlier publications [31–33]. As shown in the schematics (Extended data Fig. 3) in reference [31], the beam of light is routed by a fiber circulator to a fiber polarizer. After the polarizer the polarization of the beam is at 45° to the axis of a fiber-coupled electro-optic modulator (EOM), which generates 4.6 MHz time-varying phase shifts $\phi_m \sin(\omega t)$, where the amplitude $\phi_m = 0.92$ rad between the two orthogonal polarizations that are then launched into the fast and slow axes of a polarization maintaining (PM) single-mode fiber. Upon exiting the fiber, the two orthogonally polarized linearly polarized beams are converted into right- and left-circularly polarizations by a quarter-wave plate (QWP) and are then focused onto the sample. After reflection from the sample, the same QWP converts the reflected beams back into linear polarizations with exchanged polarization axes. The two beams then pass through the PM fiber and EOM but with exchanged polarization modes in the fiber and the EOM. At this point, the two beams have gone through the same path but in opposite directions, except for a phase difference of $\Delta\varphi$ from reflection off the magnetic sample and another time-varying phase difference by the modulation of EOM. This nonreciprocal phase shift $\Delta\varphi$ between the two counterpropagating circularly polarized beams upon reflection from the sample is twice the Kerr rotation $\Delta\varphi = 2\theta_K$. The two beams are once again combined at the detector and interfere to produce an optical signal $P(t)$:

$$P(t) = \frac{1}{2} P[1 + \cos(\Delta\varphi + \phi_m \sin(\omega t))]$$

, where P is the returned power if the modulation by the EOM is turned off. For MOKE signals that are slower that the 4.6 MHz modulation frequency used in this experiment, we can treat $\Delta\varphi$ as a slowly time-varying quantity. And $P(t)$ can be further expanded into Fourier series with the first few orders listed below:

$$P(t)/P = \frac{1}{2} [1 + \mathrm{J}_0(2\phi_m)]$$

$$+(\sin(\Delta\varphi)\, \mathrm{J}_1(2\phi_m))\sin(\omega t)$$

$$+(\cos(\Delta\varphi)\, \mathrm{J}_2(2\phi_m))\cos(2\omega t)$$

$$+2\, \mathrm{J}_3(2\phi_m))\sin(3\omega t)$$
$$+\cdots$$

, where $J_1(2\phi_m)$ and $J_2(2\phi_m)$ are Bessel J-functions. Lock-in detection was used to measure the first three Fourier components: the average (DC) power ($P_0$), the first harmonics ($P_1$), and the second harmonics ($P_2$). And the Kerr rotation can then be extracted using the following formula:

$$\theta_K = \frac{1}{2}\,\Delta\varphi = \frac{1}{2} tan^{-1}\left[\frac{J_2(2\phi_m)P_1}{J_1(2\phi_m)P_2}\right]$$

**Density functional theory (DFT) calculations:** First-principles density functional theory (DFT) with fully relativistic ultrasoft pseudopotentials [34] were performed using Quantum ESPRESSO [35]. The exchange and correlation effects were treated within the generalized gradient approximation (GGA)[36]. The plane-wave cut-off energy of 52 Ry and a 16 × 16 × 16 k-point mesh in the irreducible Brillouin zone were used in the calculations. The experimentally measured in-plane lattice parameter $a = 3.899$ Å of $Mn_3NiN$ was used in the calculations. Spin-orbit coupling and noncollinear $\Gamma_{4g}$ antiferromagnetism of $Mn_3NiN$ were included in the calculations. The anomalous Hall conductivity (AHC) was calculated using the PAOFLOW code [37]based on pseudo-atomic orbitals (PAO) [38,39].The adaptive broadening method with a 48 × 48 × 48 k-point mesh was used in the calculations.

**Atomistic spin model simulations:** Atomistic spin dynamics simulations of $Mn_3NiN$ are performed using VAMPIRE package[40] utilizing its adaptive Monte-Carlo based integrator[41] with temperature-dependent magnetization dynamics[42] using the following Heisenberg Hamiltonian,

$$H = -\frac{J_1}{2}\sum_{<ij>} \boldsymbol{S}_i.\boldsymbol{S}_j - \frac{J_2}{2}\sum_{\ll ij \gg} \boldsymbol{S}_i.\boldsymbol{S}_j - k\sum_i (\boldsymbol{S}_i.\boldsymbol{e}_i)^2$$

Here, $\boldsymbol{S}_i$ is the unit vector of the spin moment at site $i$, $J_{1(2)}$ is the magnetic exchange interaction energy between first (second) nearest neighbors, with $< ij >$ ($\ll ij \gg$) indicating a summation over all first (second) nearest neighbors, $k$ is the single ion anisotropy constant, and $\boldsymbol{e_i}$ is a unit vector normal to the cubic face where each site resides. The values used for the magnetic exchange parameters $J_1$, $J_2$, and anisotropy constant $k$ are found to be -23.9 $meV$, 24.7 $meV$, and 0.133 $meV$ (see Supplementary Note 2 for details). The time step was set to be 1fs, and periodic boundary conditions along *x, y* and *z* directions were employed for all atomistic simulations. For Néel temperature simulations, a system of 12x12x12 unit cells was used with 100,000 time steps and temperature increment of 10K. For net moment isothermal hysteresis loops, a system of 25 x 25 x 25 unit cells were subject to FC +0.3T with temperature going from 300K to 0K to obtain the appropriate domain. Then the system was heated up to the required temperature for each loop to calculate the hysteresis loop with 1,000,000 time steps at each magnetic field with an increment of 25mT going from 8T to -8T, and back to 8T.

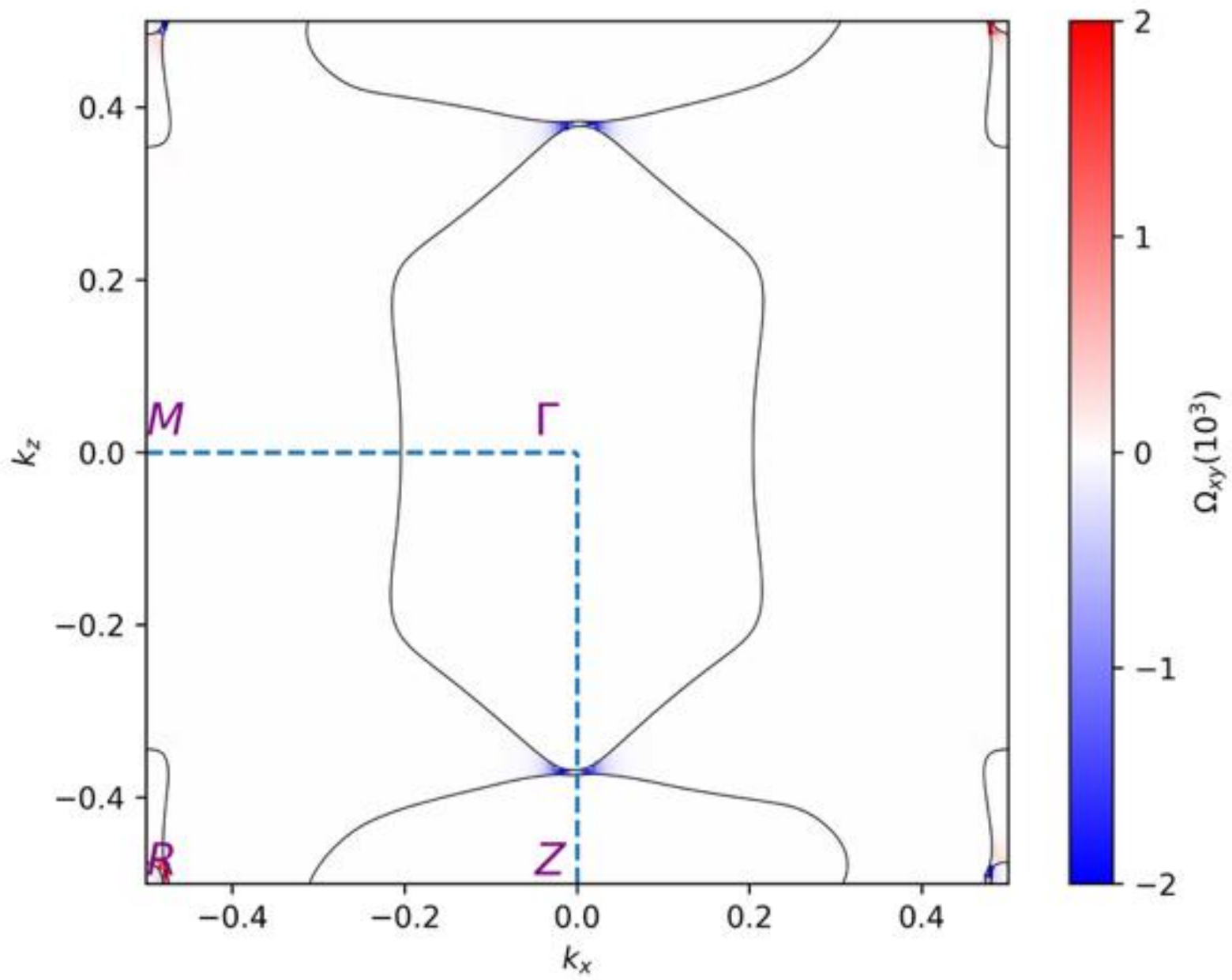

**Fig. S1. First-principles calculation of the Berry curvature of $Mn_3NiN$**. The calculated color map of the Berry curvature in the (110) plane for the $\Gamma_{4g}$ phase of $Mn_3NiN$.

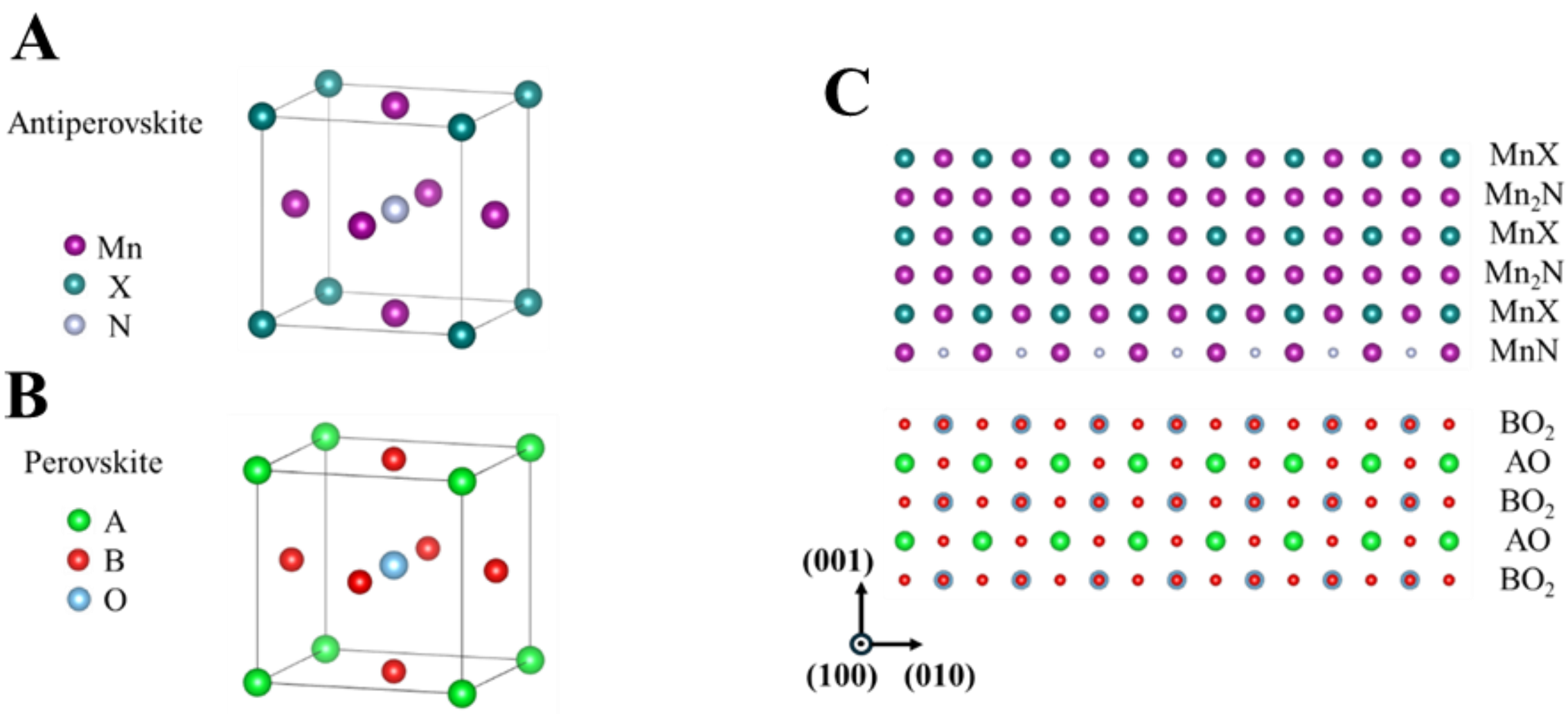

**Fig. S2. Antiperovskite and perovskite heterostructure**. Schematic illustrations of **(A)** antiperovskite nitride unit cell, **(B)** perovskite oxide unit cell. **(C)** Stacking of antiperovskite nitride on perovskite oxide in thin film heterostructure.

## Supplementary text 2

### Structural characterization of $Mn_3NiN$ thin films

A strong $Mn_3NiN$ peak is observed along with distinct Kiessig fringes, indicating high crystalline quality and a pristine interface with the substrate, as shown in fig. S3A and fig. S4 for out of plane scan around (002) peak and full range scan. This is further corroborated by the streaky RHEED pattern after growth and X-ray reflectivity data as shown in the inset to fig. S3A and fig. S5B. The narrow 0.029° full width at half maximum (FWHM) of the rocking curve for $Mn_3NiN$ (002) (fig. S5A) further supports high crystalline quality. The phi scans on the asymmetric ($\bar{1}13$) peak of $Mn_3NiN$ and LSAT, shown in fig. S6, indicate the in-plane cube-on-cube epitaxial relationship. X-ray reciprocal space mapping (RSM) measurements centered on the asymmetrical ($\bar{1}13$) peak, shown in fig. S3B, indicate that $c/a$=0.9952 at room-temperature. Atomic force microscope imaging shows an atomically smooth surface with a roughness of ~0.3 nm (fig. S3C). All these data strongly indicate high-quality epitaxial growth of $Mn_3NiN$ thin films, for which magnetic and transport measurements can probe fundamental properties.

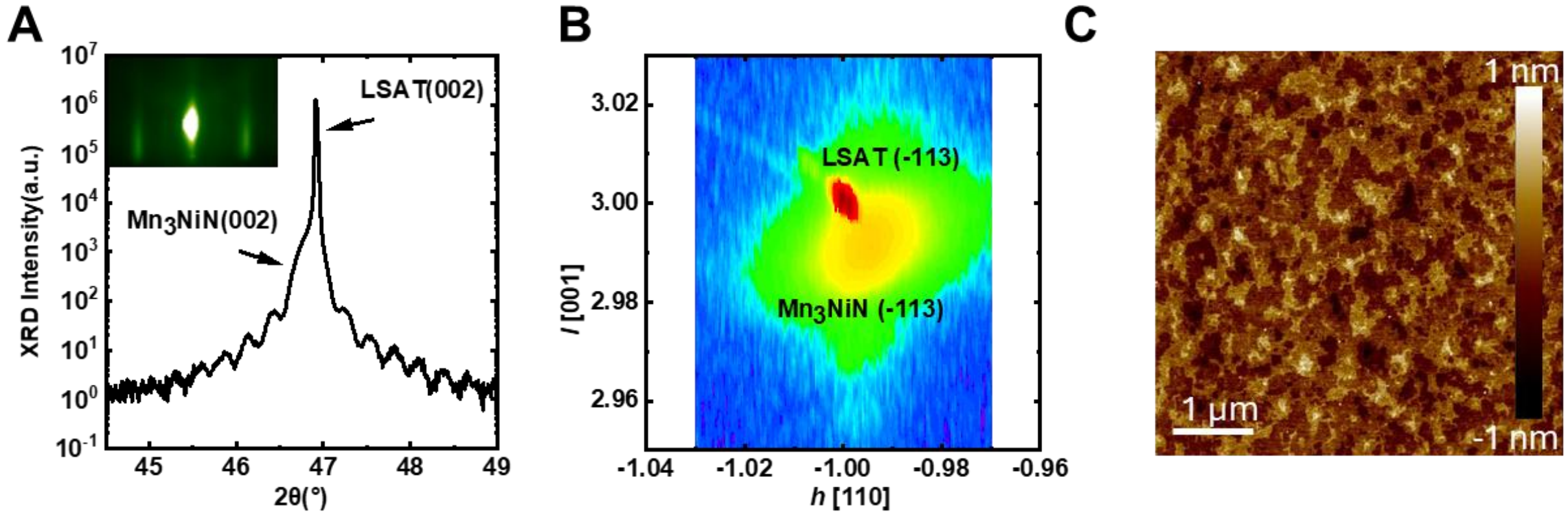

**Fig. S3. Structural and topographical characterization. (A)** θ-2θ spectrum around (002) peak of the LSAT substrate and the $Mn_3NiN$ film with the streaky RHEED pattern as inset, demonstrating that the high-quality growth of film with (001)-oriented and single phase. **(B)** X-ray reciprocal space mapping shows tetragonal distortion of $Mn_3NiN$ thin film. **(C)** atomic force imaging of $Mn_3NiN$ thin film, with the atomic level smoothness.

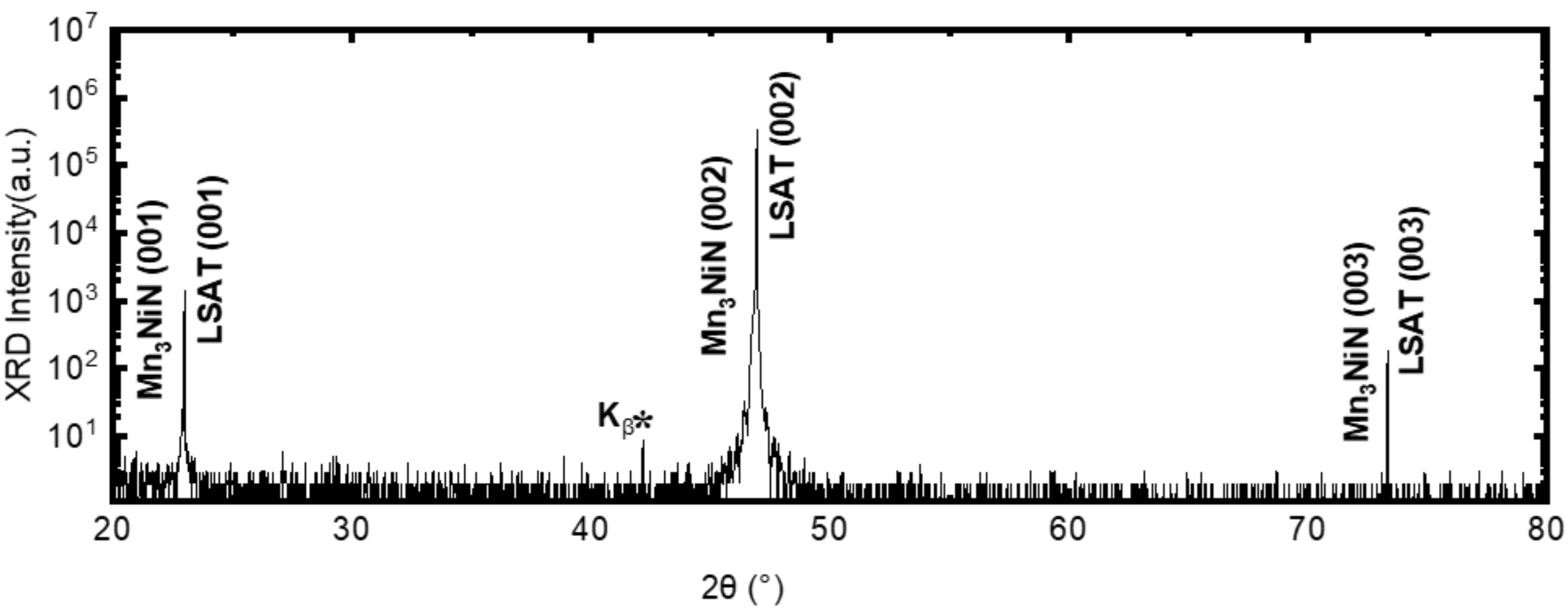

**Fig. S4. Structural characterization of Mn$_3$NiN thin films.** Wide-angle θ-2θ spectrum only shows the (00l) reflections of the LSAT substrate and the Mn$_3$NiN film, demonstrating that the film is (001)-oriented and single phase.

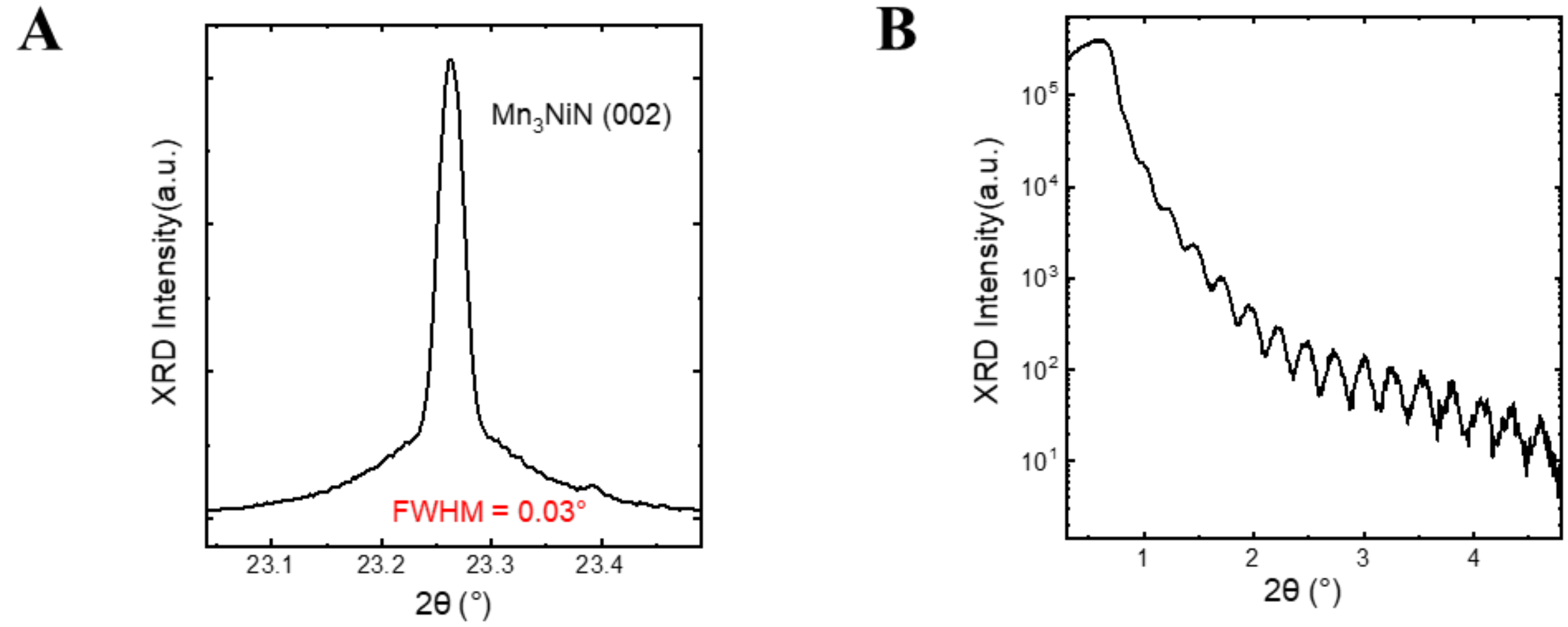

**Fig. S5. X-ray reflectivity and rocking curve analyses. (A)** X-ray reflectivity spectrum showing a pristine interface with clear fringes. **(B)** Rocking curve of (002) Mn$_3$NiN peak indicating high crystallinity with FWHM of 0.029º.

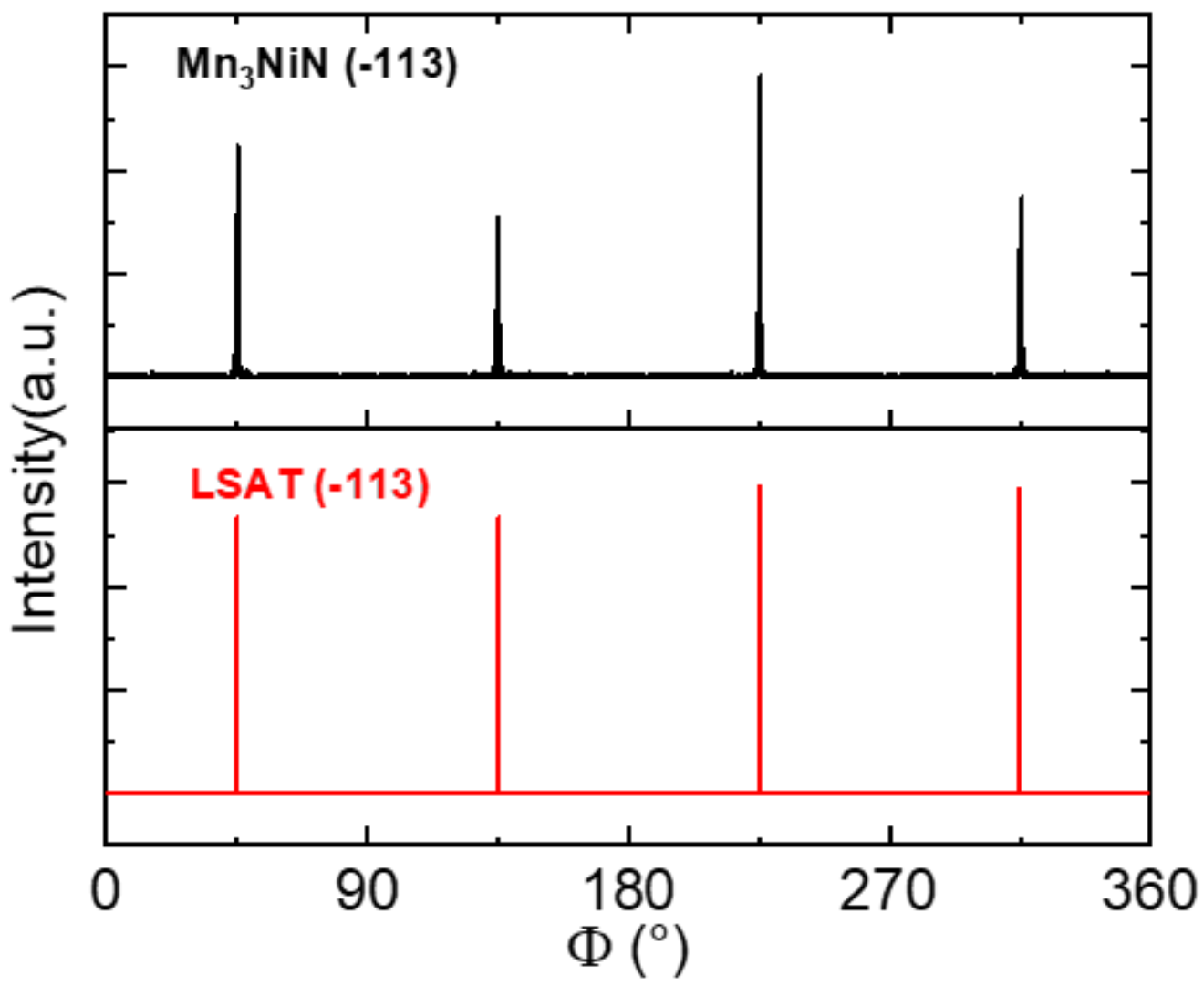

**Fig. S6. Phi-scans.** Phi-scans at asymmetric (-113) peak of $Mn_3NiN$ and LSAT indicates in-plane cube-on-cube epitaxial relationship.

## Supplementary text 3

### Magnetic characterization of $Mn_3NiN$ thin films

Fig. S7A shows temperature dependence of the out-of-plane magnetic moment. From the variation of magnetization with temperature, we clearly note a deviation between field-cooling (FC) and zero-field-cooling (ZFC) curves below the Néel temperature of 240 K (note fig. S9A for in-plane magnetic moment). A field-cooled magnetic moment of ~8 m$\mu_B$/Mn indicates the presence of a weak ferromagnetic moment, like that observed in other non-collinear AFM materials. Isothermal magnetic field sweeps above and below the Néel temperature (fig. S7B) show hysteretic behavior at 230K (below $T_N$) with a weak saturation moment consistent with the field-cooled measurement of fig. S7A, and nonhysteretic behavior at 250K (above $T_N$). At 230K the applied out-of-plane magnetic field switches the weak net moment with a coercive field of a 0.4 T and a saturation field of several Tesla. The relatively large coercive field suggests strong crystalline anisotropy. The ~8 m$\mu_B$/Mn saturation moment is much smaller than the 2.7 $\mu_B$/Mn moment that would arise from fully-aligned Mn spins in this structure. We further investigated temperature dependent neutron diffraction measurements on these high-quality $Mn_3NiN$ thin films (fig. S7C and fig. S8). Interestingly, the $Mn_3NiN$ (001) magnetic diffraction peak was observed below $T_N$, consistent with the onset of long-range AFM $\Gamma_{4g}$ spin ordering. The magnetic peak intensity remains almost constant down to 2 K (fig. S8). Thus, we conclude the presence of robust $\Gamma_{4g}$ phase in these $Mn_3NiN$ films.

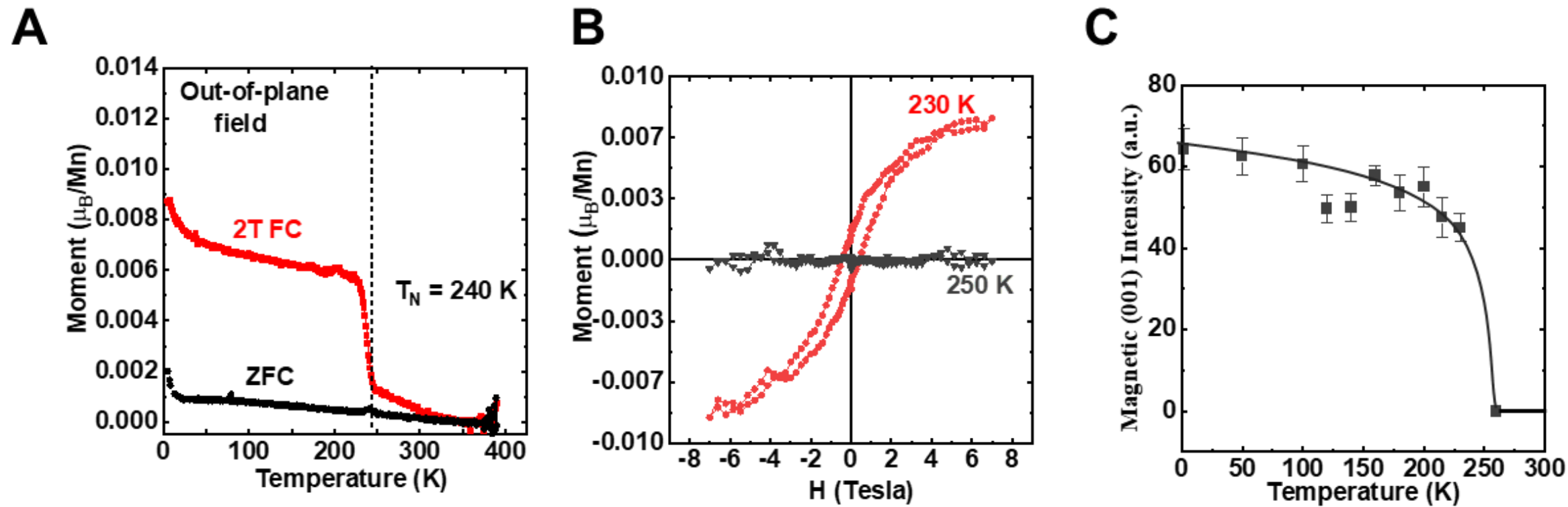

**Fig. S7. Magnetic characterization of $Mn_3NiN$ thin films. (A)** Out-of-plane net magnetization vs temperature curves for 2 T field cooled (red) and zero-field-cooled (black), while measuring during warming in 0.5 T. Distinct Néel transitions are visible around 240 K. **(B)** Net magnetization vs out-of-plane magnetic field showing clear weak net moment below $T_N$. **(C)** Neutron diffraction measurement on a 250nm $Mn_3NiN$ films, temperature dependent neutron diffraction intensity indicate the $\Gamma^{4g}$ phase in $Mn_3NiN$.

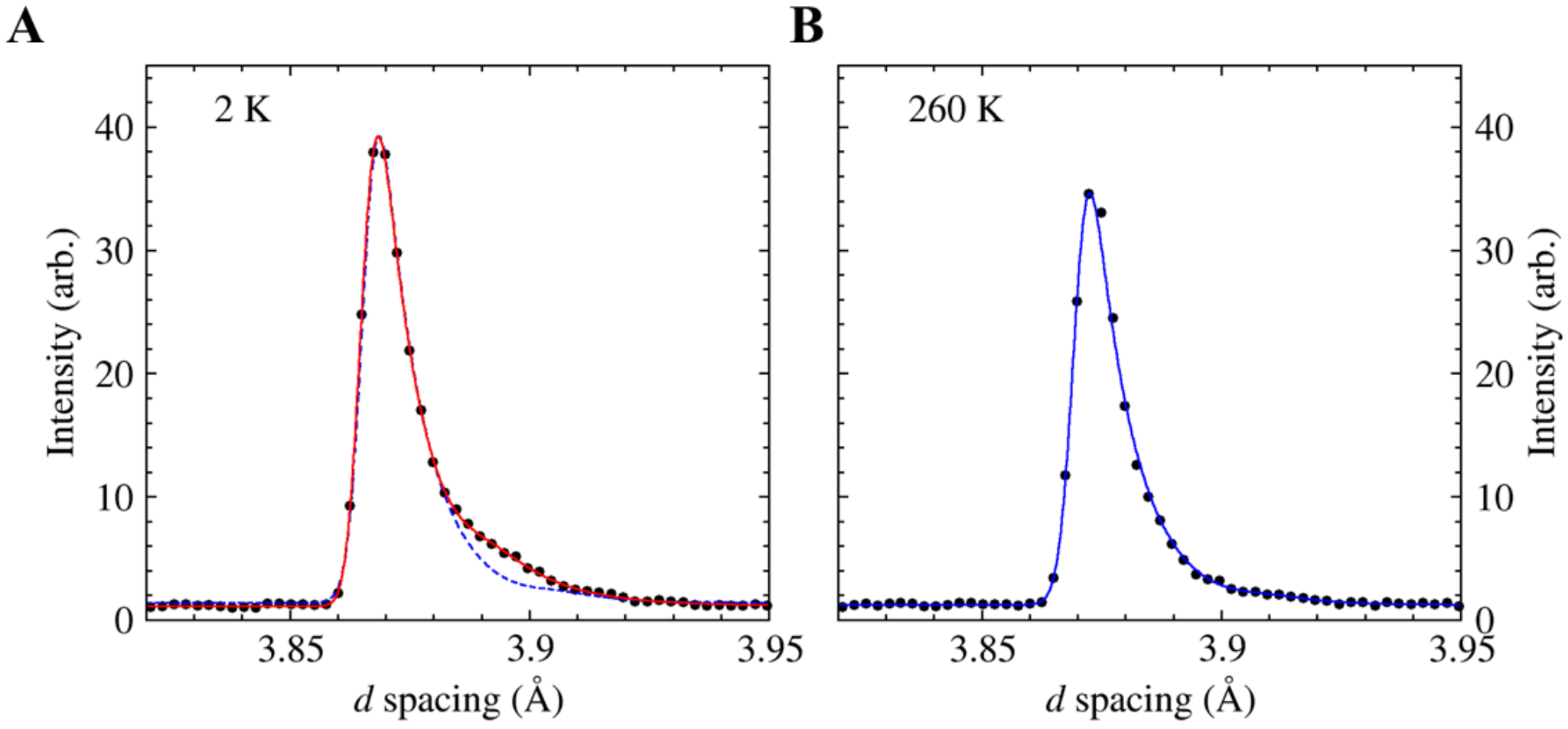

**Fig. S8. Temperature dependent neutron diffraction data.** A single 250 nm $Mn_3NiN$ thin film sample was mounted within a $^4$He cryostat and oriented for scattering in the HHL reciprocal space plane in a geometry optimized for peak flux and resolution at the (001) Bragg reflection. Neutron diffraction intensity (black points) at the (001) reciprocal space position defined with respect to the antiperovskite crystal structure, measured at **(A)** 2 K and **(B)** 260 K. The 2 K intensity includes contributions from both nuclear diffraction of the substrate and magnetic diffraction of the film, while the 260 K data include only diffraction from the substrate. The 2 K data have been fit using two back-to-back exponential peak profiles (red line), and the 260 K data have been fit with a single back-to-back exponential peak profile (blue line). The single 260 K peak profile has been overlaid in pane **(A)**, shifted and scaled to the more intense substrate peak (blue dashed line), revealing the weak magnetic intensity from the film at longer *d*-spacing.

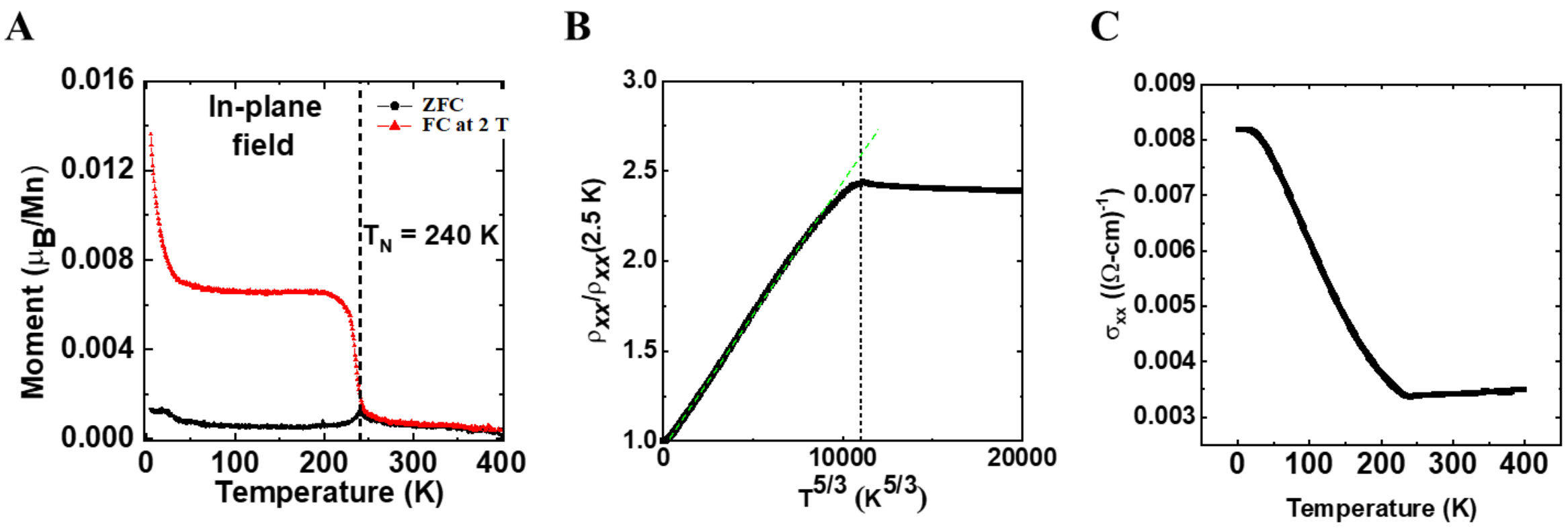

**Fig. S9. In-plane magnetization data and residual resistivity plot.** **(A)** In-plane net magnetization vs temperature curves for field cooled (red) and zero-field-cooled (black). **(B)** Temperature dependent variation of sheet resistivity and **(C)** Temperature dependent variation of sheet conductivity showing excellent metallic behavior below $T_N$ as of 240K.

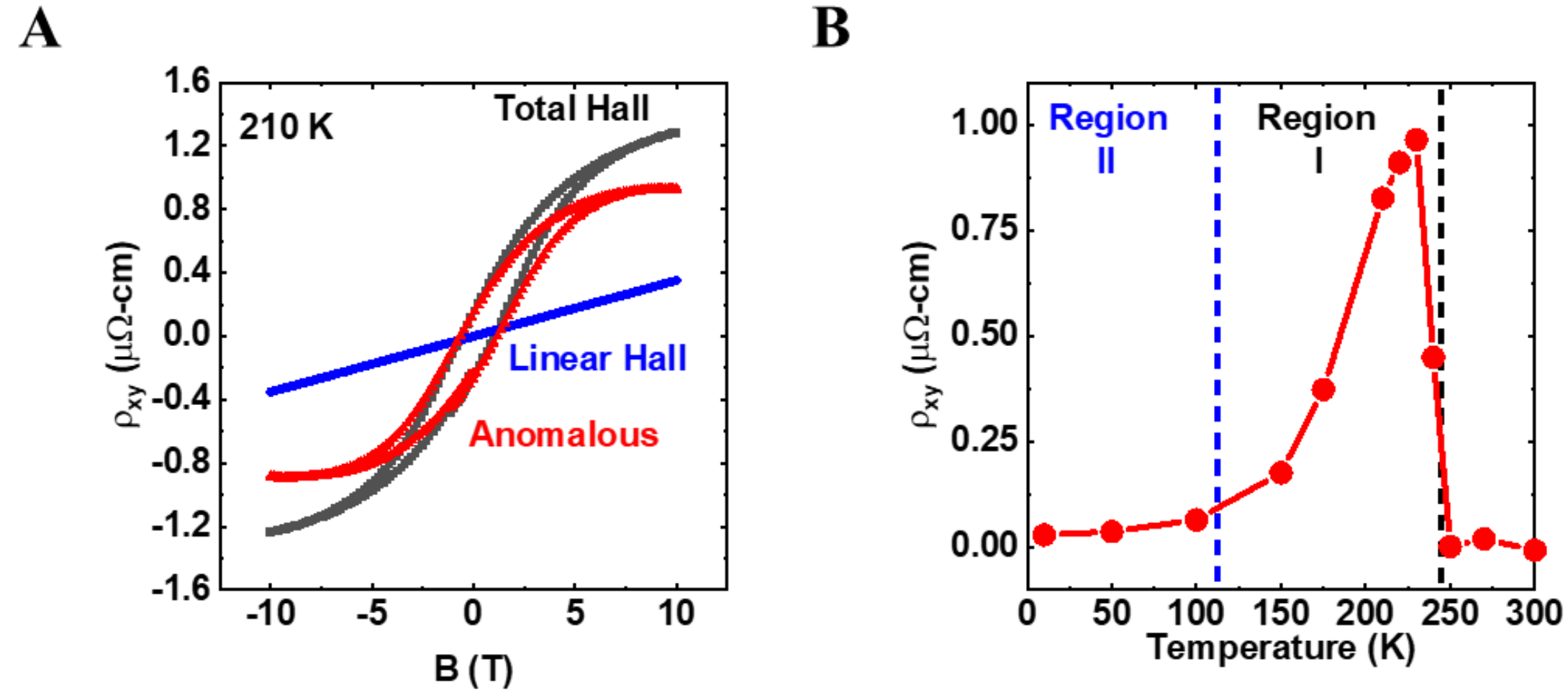

**Fig. S10. Magneto transport measurements in a Hall-bar device.** **(A)** Hall resistivity vs out of plane magnetic field at 210 K, which also shows both normal linear part and anomalous part separated from the total. **(B)** Temperature-dependent anomalous Hall resistivity.

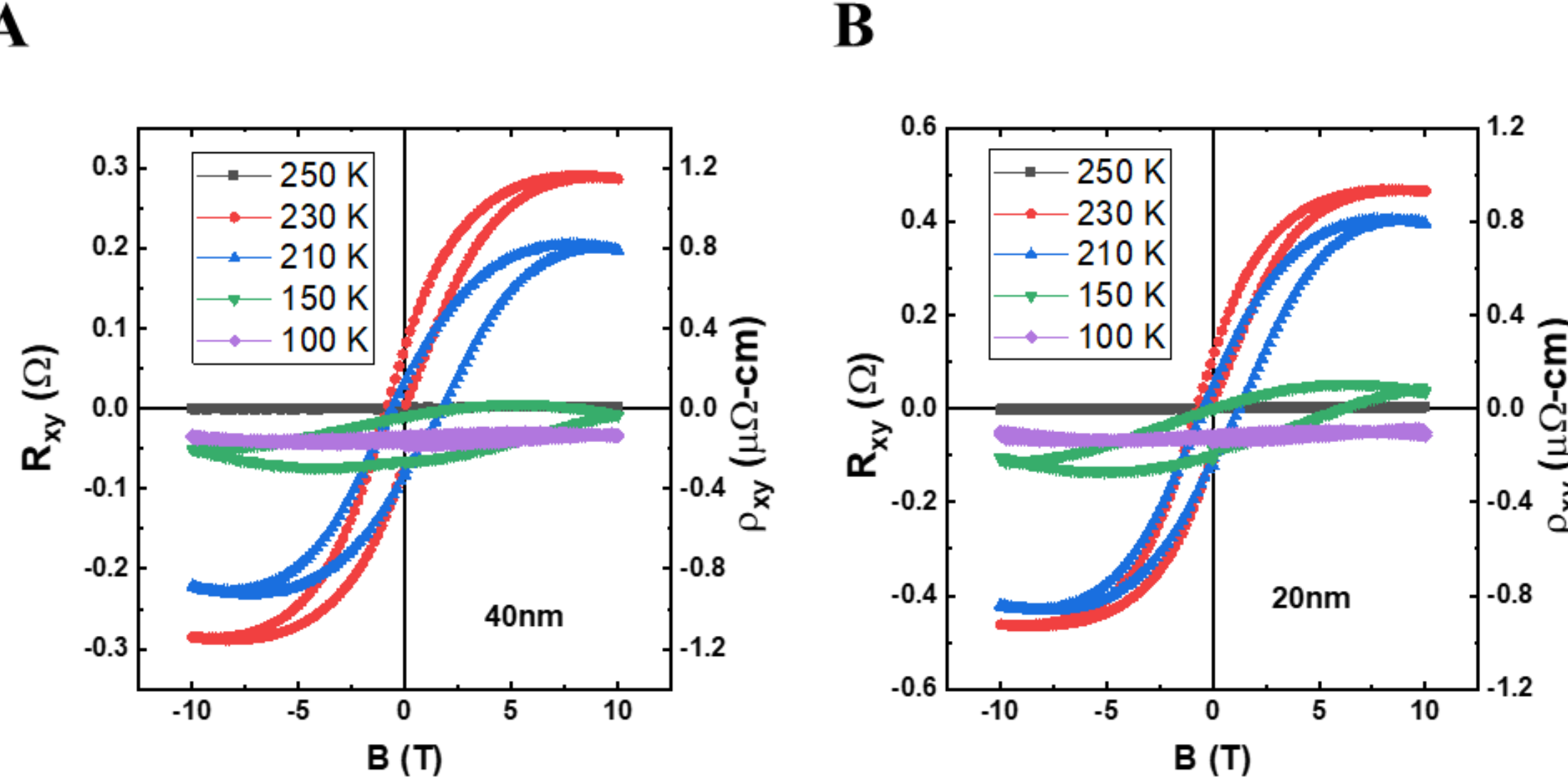

**Fig. S11. Thickness dependent of AHE in $Mn_3NiN$. (A)** Temperature dependent of anomalous Hall resistance and resistivity of 40nm and **(B)** 20nm $Mn_3NiN$ single-crystal thin-films.

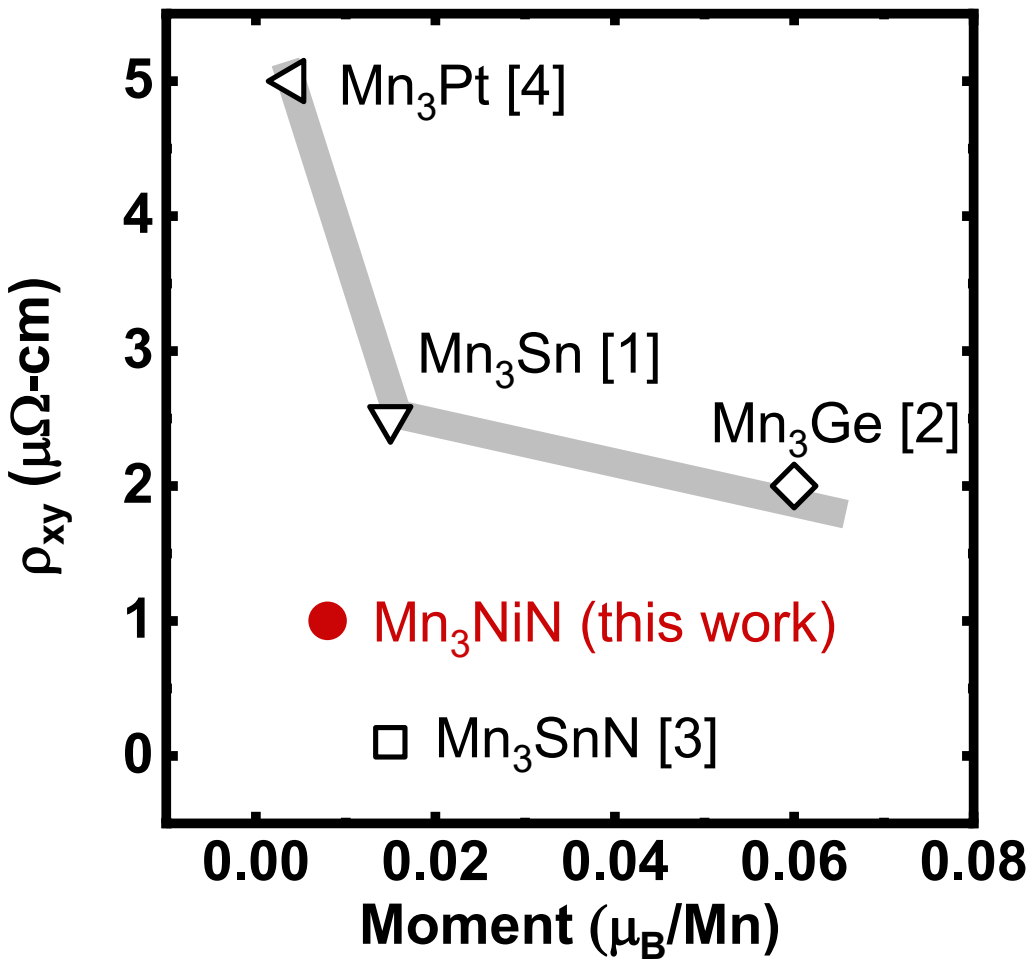

**Fig. S12. Comparison of AHE with other non-collinear antiferromagnetic materials** Comparison of AHE and net moment of $Mn_3NiN$ in this work with other non-collinear antiferromagnetic materials with the reference [1-4].

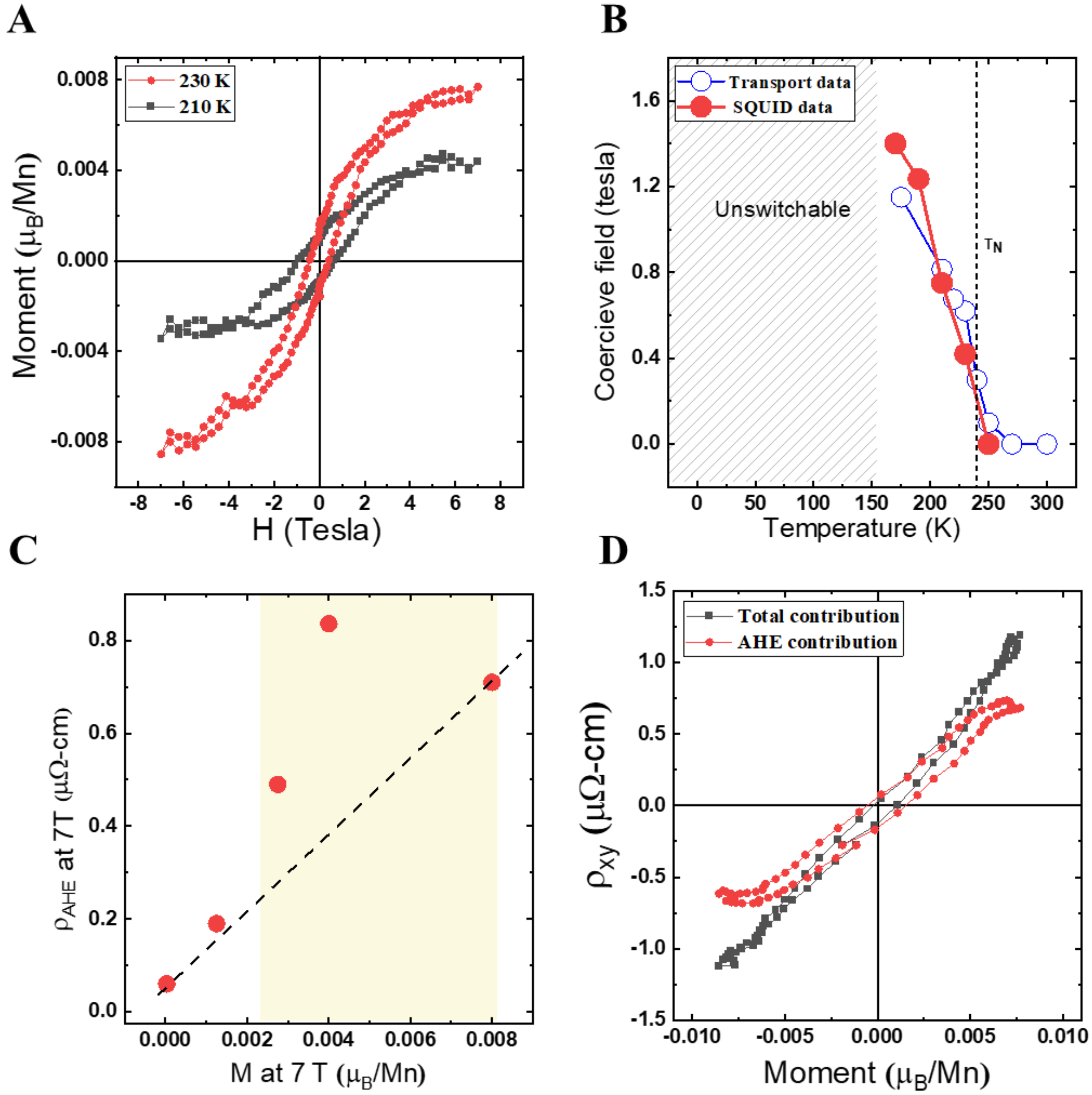

**Fig. S13. Temperature dependent variations of coercive field and remanent moment. (A)** Isothermal M vs H measurements for out-of-plane applied magnetic field for $Mn_3NiN$ thin-films. **(B)** Variation of extracted coercive field with temperature. Increasing Coercive field below $T_N$, indicates enhanced magnetic domain anisotropy. **(C)** Variation of extracted ρAHE at 7 T with the corresponding magnetization obtained at various temperatures shows clear anomalous behavior. **(D)** Variation of ρxy at 230 K with the corresponding magnetization data, which shows deviation indicating different origin of AHE than simple magnetism.

**Supplementary text 4**

**Sagnac characterization of $Mn_3NiN$ thin films**

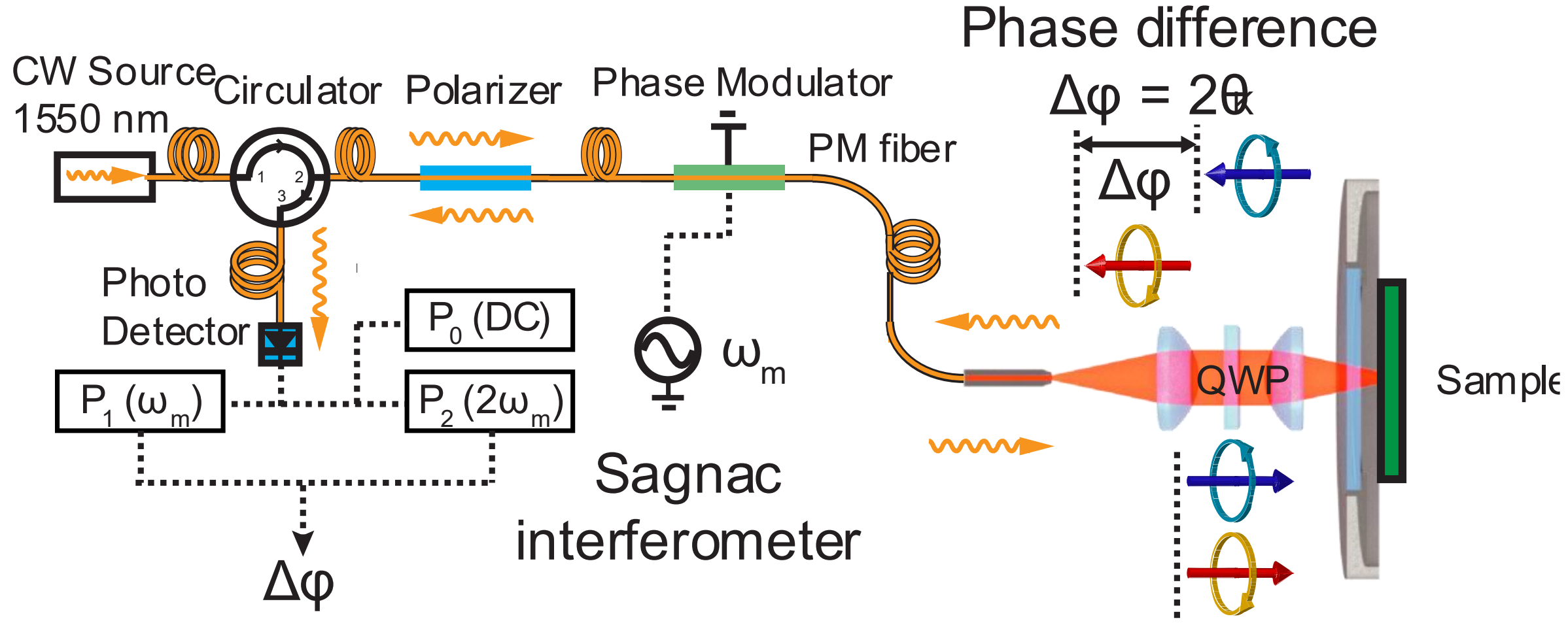

**Fig. S14. Schematic drawing of the basic setup of Sagnac MOKE interferometry.** MOKE signal $\theta_K$ is measured by a Sagnac zero-area-loop fiber-optic interferometer during varying temperature and sweeping magnetic field.

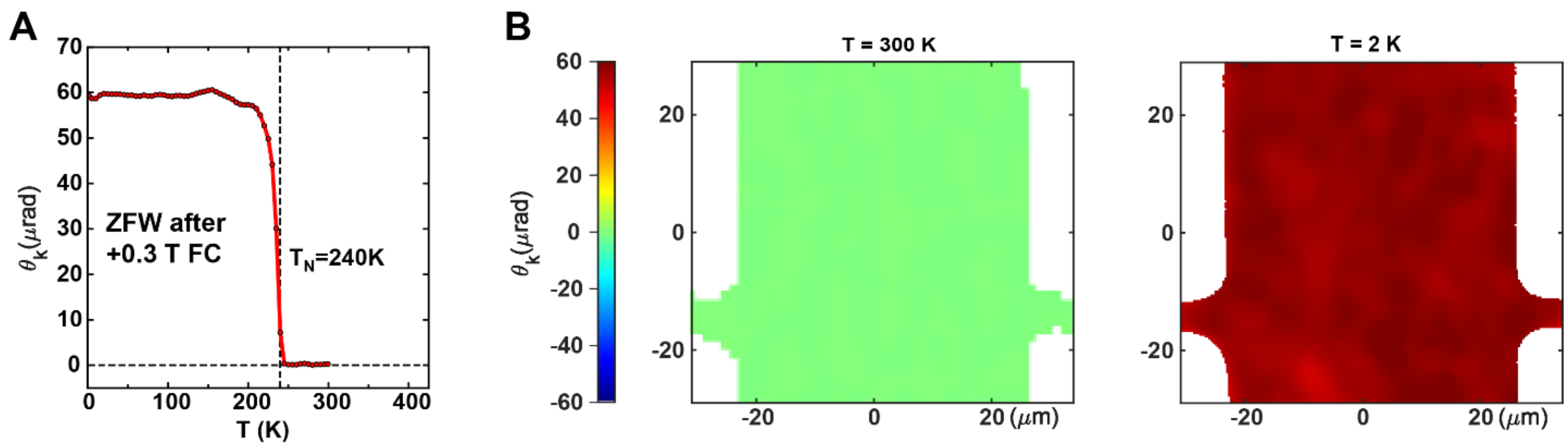

**Fig. S15. Temperature dependent Sagnac MOKE characterization of $Mn_3NiN$ thin films. (A)** Temperature dependence of the Kerr rotation $\theta_k$ measured during zero field warmup after +0.3 T field cooling (FC), showing an onset of $\theta_k \approx 60\ \mu rad$ below the Néel temperature $T_N \approx 240$ K. **(B)** $\theta_k$ imaging at 300 K and 2 K, before and after warming in zero field following a +0.3 T FC, demonstrating the emergence of a strong Kerr signal below $T_N$.

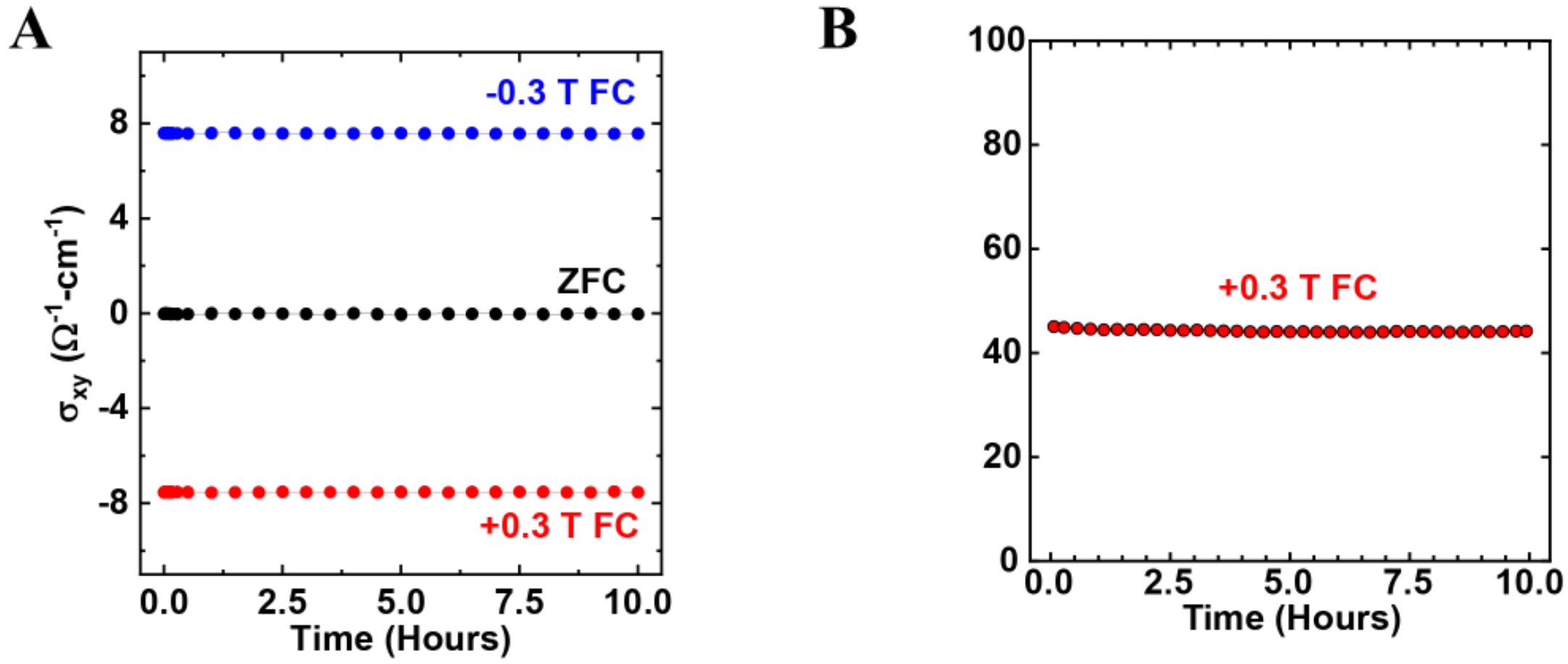

**Fig. S16. Time dependent AHE and Kerr signal.** Time dependent **(A)** anomalous Hall conductivity and **(B)** Kerr signal under +/- 0.3T and ZFC, showing the robust stability with the potential for permanent memory.

## Supplementary text 5

### Parameters for atomistic spin model simulations

The first nearest neighbor Heisenberg exchange energy $J_1$ is calculated via DFT by comparing the energy of the $\Gamma_{4g}$ (fig. S17A) and ferromagnetic (fig. S17C) phases, as follows:

$$E^{\Gamma 4g} = E_0 + 6J_1 - 18J_2$$
$$E^{\mathrm{FM}} = E_0 - 12J_1 - 18J_2$$
$$J_1 = \frac{E^{\Gamma 4g} - E^{\mathrm{FM}}}{18} = -23.85\ meV$$

The second nearest neighbor exchange energy $J_2$ is obtained by varying to fit experimentally measured $T_N$ (fig. S17D).

The single ion magnetic anisotropy form respects the Mn atom 4/mmm site symmetry of the Wykoff position 3c in its space group Pm-3m. The positive (negative) sign of the anisotropy constant $k$ results in the noncollinear antiferromagnetic phase $\Gamma_{4g(5g)}$, while increasing the magnitude of $k$ monotonically increases the net moment. It is calculated via DFT by comparing the energy of the $\Gamma_{4g}$ (fig. S17A) and $\Gamma_{5g}$ (fig. S17B) phases, as follows:

$$E^{\Gamma 4g} = E'_0 - 3k\ \left(\frac{2}{\sqrt{6}}\right)^2$$
$$E^{\Gamma 5g} = E'_0 - 3k\ (0)^2$$
$$k = \frac{E^{\Gamma 4g} - E^{\Gamma 5g}}{2} = -0.133\ meV$$

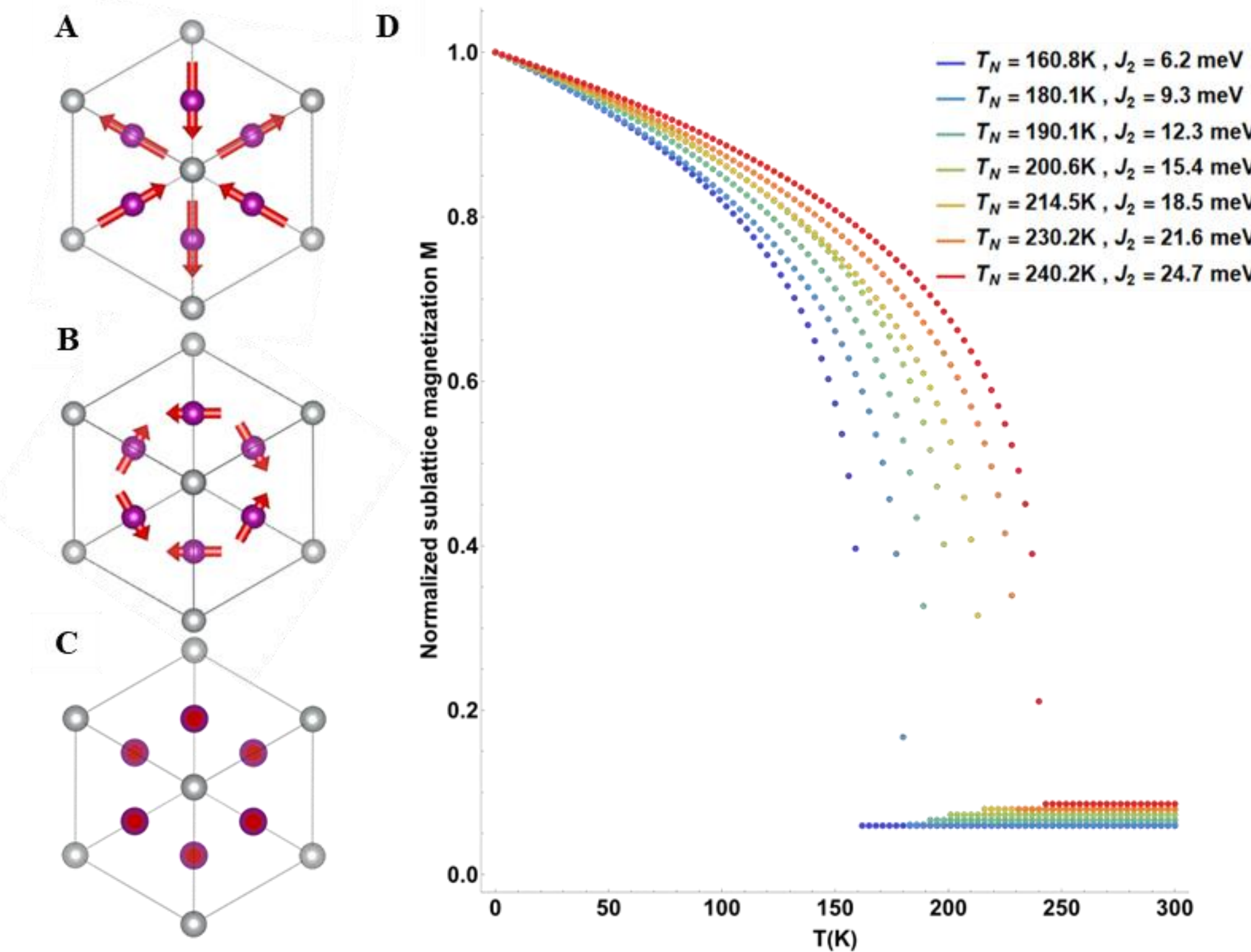

**Fig. S17. Parameters for atomistic spin model simulations.** Schematics of the $Mn_3NiN$ magnetic unit cell projected to the (111) plane for noncollinear antiferromagnetic **(A)** $\Gamma_{4g}$ and **(B)** $\Gamma_{5g}$ phases, and **(C)** ferromagnetic phase. **(D)** Néel temperature $T_N$ simulations for different second nearest neighbor exchange energy $J_2$.